\documentclass[english]{emulateapj}

\usepackage{graphicx}
\usepackage{amssymb}
\usepackage{natbib}
\usepackage{amsmath}

\makeatletter
\providecommand{\tabularnewline}{\\}

\providecommand{\boldsymbol}[1]{\mbox{\boldmath $#1$}}
\newcommand{\kmps}{\mathrm{km~s^{-1}}}
\newcommand{\Kelvin}{\mathrm{K}}
\newcommand{\Msun}{\mathrm{M_{\sun}}}

\newcommand{\Lsun}{L_{\sun}}
\newcommand{\MsunPerYear}{\mathrm{M_{\sun}\,yr^{-1}}}

\makeatother

\slugcomment{}
\shorttitle{}
\shortauthors{3-D simulations of Precessing Outflows from AGN}

\begin{document}

\title{Three-Dimensional Simulations of Inflows Irradiated by a Precessing
Accretion Disk in Active Galactic Nuclei: Formation of Outflows}

\author{Ryuichi Kurosawa and Daniel Proga}

\affil{Department of Physics and Astronomy, University of Nevada Las Vegas,
Box~454002, 4505~Maryland Pkwy, Las Vegas, NV 891541-4002}

\email{\{rk,dproga\}@physics.unlv.edu}

\begin{abstract}
We present three-dimensional (3-D) hydrodynamical simulations of gas
flows in the vicinity of an active galactic nucleus (AGN) powered
by a precessing accretion disk. We consider the effects of the radiation
force from such a disk on its environment on a relatively large scale
(up to $\sim10$~pc). We implicitly include the precessing disk by
forcing the disk radiation field to precess around a symmetry axis
with a given period ($P$) and a tilt angle ($\Theta$). We study time
evolution of the flows irradiated by the disk, and investigate basic
dependencies of the flow morphology, mass flux,
angular momentum on different combinations of $\Theta$ and $P$. As
this is our first attempt to model such 3-D gas flows, we consider
a simplest form of radiation force i.e., force due to electron scattering,
and neglect the forces due to line and dust scattering/absorption.
Further, the gas is assumed to be nearly isothermal. 
We find the gas flow settles into a configuration with two components,
(1) an equatorial inflow and (2) a bipolar inflow/outflow with the
outflow leaving the system along the poles (the directions of disk
normals). However, the flow does not always reach a steady state.
We find that
the maximum outflow velocity and the kinetic outflow power at the
outer boundary can be reduced significantly with increasing $\Theta$.
We also find that of the mass inflow rate across the inner boundary
does not change significantly with increasing $\Theta$. The amount
of the density-weighted mean specific angular momentum deposited to
the environment by the precessing disk increases as $P$ approaches
to the gas free-fall time ($t_{\mathrm{ff}}$), and then decreases
as $P$ becomes much larger than $t_{\mathrm{ff}}$. Generally, the
characteristics of the flows are closely related to a combination
of $P$ and $\Theta$, but not to $P$ and $\Theta$ individually. Our
models exhibit helical structures in the weakly collimated outflows.
Although on different scales, the model reproduces the Z- or S- shaped
density morphology of gas outflows which are often seen in radio observations
of AGNs.
\end{abstract}


\section{Introduction}

\label{sec:Introduction}

Powered by accretion of matter onto a super massive ($10^{6}$--$10^{10}\,\Msun$)
black hole (SMBH), Active Galactic Nuclei (AGN) release large amount
of energy (e.g., \citealt{Lynden-Bell:1969}) as electromagnetic radiation
($10^{10}$--$10^{14}\Lsun$) over a wide range of wavelengths, from
the X-ray to the radio. The very central location of AGN in their
host galaxies indicates that the radiation from AGN can play an important
role in determining the physical characteristics (e.g., the ionization
structure, the gas dynamics, and the density distribution) of their
surrounding environment in different scales i.e., from the scale of
AGN itself to a lager galactic scale, and even to an inter-galactic
scale (e.g., \citealt{Quilis:2001}; \citealt{DallaVecchia:2004};
\citealt{McNamara:2005}; \citealt{Zanni:2005}; \citealt{Fabian:2006};
\citealt{Vernaleo:2006}). The feedback process of AGN in the form
of mass or energy outflows, in turn, is one of key elements in
galaxy formation/evolutionary models (e.g., \citealt{ciotti:1997},
\citeyear{ciotti:2001}, \citeyear{ciotti:2007}; \citealt{Silk:1998}; \citealt{king:2003};
\citealt{Begelman:2005}; \citealt{Hopkins:2005}; \citealt{Murray:2005}; \citealt{Sazonov:2005};
\citealt{Silk:2005};
\citealt{Springel:2005}; \citealt{Brighenti:2006}; \citealt{Fabian:2006b}; \citealt{Fontanot:2006};
\citealt{Thacker:2006}; \citealt{Wang:2006}, \citealt{Tremonti:2007}). 

Although the AGN outflows can be driven by magnetocentrifugal force
(e.g., \citealt{Blandford:1982}; \citealt{Emmering:1992};
\citealt{Konigl:1994}; \citealt{Bottorff:1997}) and thermal pressure
(e.g., \citealt{Weymann:1982}; \citealt{Begelman:1991};
\citealt{Everett:2007}), it is the radiation force from the luminous
accretion disk that is most likely the dominant force driving winds
capable of explaining the blueshifted absorption line features often
seen in the UV and optical spectra of AGN
(e.g.,~\citealt{Shlosman:1985}; \citealt{Murray:1995b}; \citealt{Proga:2000};
\citealt{Proga:2004}). In reality, these three forces may interplay and
contribute to the dynamics of the outflows in AGN in somewhat
different degrees.

Another complication in the outflow gas dynamics is the presence of
dust. The radiation pressure on dust can drive dust outflows,
and their dynamics is likely to be coupled with the gas dynamics (e.g.,
\citealt{Phinney:1989}; \citealt{Pier:1992}; \citealt{Emmering:1992};
\citealt{Laor:1993}; \citealt{Konigl:1994}; \citealt{Murray:2005}).
The AGN environment on relatively large scales ($10^{2}-10^{3}$~pc)
is known to be a mixture of gas and dust (e.g.~\citealt{Antonucci:1984};
\citealt{Miller:1990}; \citealt{Awaki:1991}; \citealt{Blanco:1990};
\citealt{Krolik:1999}); however, in much smaller scales ($<\sim10$~pc)
one does not expect much dust to be present because the temperature
of the environment is high ($>10^{4}\,\Kelvin$). Concentrating on
only the gas component, the dynamics of the outflows in smaller scales
was studied by e.g..~\citet{Arav:1994}, \citet{Proga:2000} in
1-D and 2-D, respectively. 

Radio observations show that a significant fraction of extended extragalactic
sources display bending or twisting jets from their host galaxies.
For example, \citet{Florido:1990} found that $\sim11$\% of their
sample (368 objects) show anti-symmetrically bending jets (S-shaped
or Z-shaped morphology) while $\sim9$\% show the symmetrically bending
jets (U-shaped morphology). Similarly \citet{Hutchings:1988} studied
the morphology of the radio lobes from 128 quasars (with $z<1$),
and found that $30$\% of the sample show a sign of bending jets.
The bending and misalignment of jets are also observed in parsec scales
in compact radio sources (e.g.~\citealt{Linfield:1981}; \citealt{Appl:1996};
\citealt{Zensus:1997}). Examples of the radio maps displaying the
S- or Z-shaped morphology of jets can be found in e.g.~\citet{Condon:1984},
\citet{Hunstead:1984}, and \citet{Tremblay:2006}. 

Using the data available in literature, \citet{Lu:2005} compiled
the list (see their Tab.~1) of 41 known extragalactic radio sources
which show an evidence of jet precession, along with their jet precession
periods ($P$) and the half-opening angle ($\psi$) of jet precession
cones. According to this list, a large fraction (67\%) of system has
rather small half-opening angles, i.e., $\psi<\sim15^{\circ}$. A
large scatter in the precession periods are found in their sample;
however, most of the precession periods are found in between $10^{4}$
and $10^{6}$~yr (see also \citealt{Roos:1988}). Note that the precession
periods are usually too long to be determined directly by variability
observations. Typically the precession periods are found by fitting
the radio map with a kinematic jet model (e.g., \citealt{Gower:1982};
\citealt{Veilleux:1993}). Interestingly, \citet{Appl:1996} showed
that a typical precession period of tilted massive torus around SMBH
is $\sim10^{6}$~yr.

The S- and Z-shaped morphology seen in the observations mentioned
above can naturally explained by precessing jets. Further, the precessing
of jets can occur if the underlying accretion disk is tilted
(or warped) with respect to the symmetry plane. There are at least
five known mechanisms that can causes warping and precession of in
accretion disks (1)~the Bardeen-Petterson effect (\citealt{bardeen:1975};
see also \citealt{Schreier:1972}; \citealt{Nelson:2000}; \citealt{Fragile:2005};
\citealt{King:2005}), (2)~tidal interactions in binary BH system
(e.g., \citealt{Roos:1988}; \citealt{Sillanpaa:1988}; \citealt{katz:1997};
\citealt{Romero:2000}; \citealt{Caproni:2004}), (3)~radiation-driven
instability (e.g., \citealt{Petterson:1977}; \citealt{Pringle:1996};
\citealt{Maloney:1996}; \citealt{Armitage:1997}), (4)~magnetically-driven
instability (\citealt{Aly:1980}; \citealt{Lai:2003}), and (5)~Disk-ISM
interaction (e.g., \citealt{Quillen:1999}). Using a small sample
of AGN, \citet{Caproni:2006b} examined whether mechanisms (1)--(4)
are capable of explaining the observed precession periods. Similarly
\citet{Tremblay:2006} searched for a possible cause of disk precession
and warping of the FR~I radio source 3C~449 using mechanisms~(2),
(3) and (4) above. In general, it is very difficult to determine the exact cause of disk/jet
precession for a given AGN system because of large uncertainties in model
parameters and observed precession periods (which are also often model
dependent). 

\citet{Kochanek:1990} presented a hydrodynamical simulation of jet
propagation along the surface of an axisymmetric hollow/cone to approximate
a jet with fast precession; however, intrinsically non-axisymmetric
nature of the dynamics of jet precession requires the problem to be
solved/simulated in 3-D. Hydrodynamical simulations of extragalactic
radio sources with precessing jets in full 3-D have been performed
by e.g., \citet{Cox:1991}, \citet{Hardee:1992}, \citet{Hardee:1994},
Typically, in these models, the jets are driven at the origin by a
small-amplitude precession to break the symmetry and excite helical
modes of the Kelvin-Helmholtz instability. Careful stability/instability
analysis of such simulations has been presented by \citet{Hardee:1995}.
The effect of magnetic field has been also investigated by e.g., \citet{Hardee:1995b}
while the effect of optically thin radiative cooling on the Kelvin-Helmholtz
instability has been investigated by e.g., \citet{Xu:2000}. 
Precession of relativistic jets in 3-D with or without magnetic field
has been also studied (e.g., \citealt{Hardee:2001};
\citealt{Hughes:2002}; \citealt{Aloy:2003}; \citealt{Mizuno:2007}).
On much larger scales, \citet{Sternberg:2007} studied the
effect of precessing massive slow jets onto the intergalactic medium
(IGM) in a galaxy cluster, and found such jets can inflate a fat bubble
in the IGM. 
In the models mentioned above, jets themselves are injected on small
scales, and the jet propagations are studied. However, it is also
possible to model a self-consistent production of a jet and its
subsequent propagation. For example, a jet can be produced from an
infalling matter by radiation pressure due to
a luminous accretion disk (e.g., see \citealt{Proga:2007};
\citealt{Proga:2007b}, for axisymmetric cases).

Regardless of the exact cause of disk/jet precession,
the observations (e.g.~\citealt{Florido:1990}; \citealt{Hutchings:1988})
suggest that a significant fraction of AGN contain warped or precessing
disks. One might expect the details of the radiative feedback
processes in such systems are different from the ones predicted by
axi-symmetry models (e.g.~\citealt{Proga:2000}; \citealt{Proga:2007};
\citealt{Proga:2007b}). If they differ, then by 
how much? In this paper, we explore the effects of disk precession
on the gas dynamics in the AGN environment by simulating the outflows
driven by the radiation force from a luminous precessing accretion
disk around a SMBH. Specifically, we will examine how the mass-accretion
rate, the outflow powers (kinetic and thermal), the morphology of
the flows, and the specific angular momentum of the gas are affected
by the presence of a precessing disk and its radiation field. This
is our first step toward a full extension of the axisymmetric radiation-driven
wind model of \citet{Proga:2007} to a full 3-D model.

In the following section, we describe our method and model assumptions,
and we give the results of our 3-D hydrodynamical simulations in \S~\ref{sec:Results}.
Our conclusions are summarized in \S~\ref{sec:Conclusions}.

\section{Method}

\begin{figure*}
\begin{center}

\includegraphics[width=0.95\textwidth]{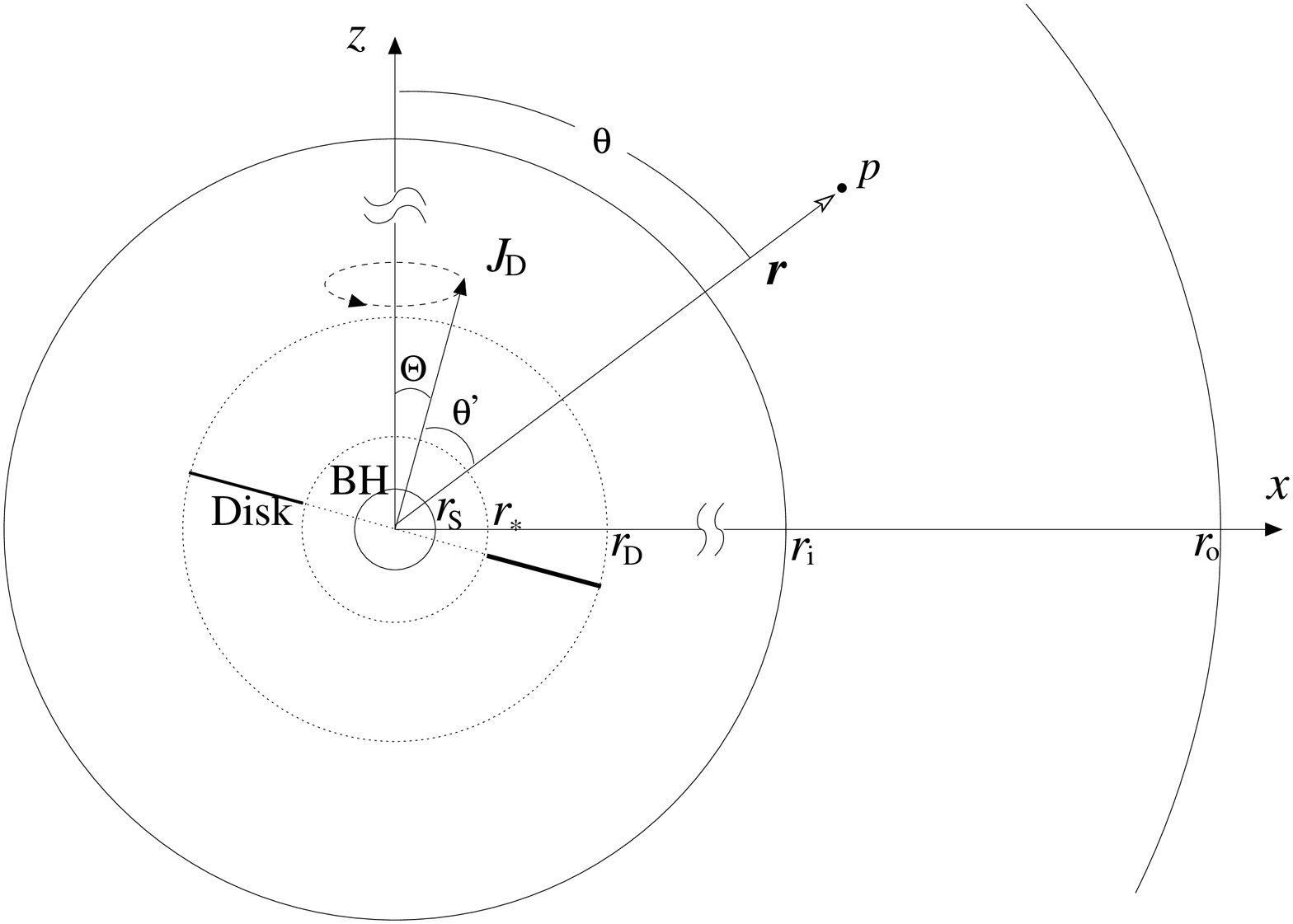}

\end{center}

\caption{Basic model configuration. A super massive blackhole (BH) with its
Schwarzschild radius $r_{S}$ is located at the center of the cartesian
coordinate system ($x$, $y$, $z$) where the $y$-axis is perpendicular
to and into the page. The normal vector or the angular momentum ($\boldsymbol{J}_{\mathrm{D}}$)
of the accretion disk, spanning from its inner radius $r_{*}$ to
its outer radius $r_{D}$, is misaligned with the $z$-axis by a tilt
angle $\Theta$ i.e., the accretion disk is tilted by $\Theta$ from
the $x$-$y$ plane. The accretion disk hence
its angular momentum $\boldsymbol{J}_{\mathrm{D}}$ is assumed to
precess around the $z$-axis with the precession period $P$. The
3-D hydrodynamic simulations are performed in the spherical coordinate
system ($r$, $\theta$, $\phi$). The simulations are performed in the
radial range between the inner boundary
$r_{\mathrm{i}}$ and the outer boundary $r_{\mathrm{o}}$. The
radiation pressure from the central BH on a point $p$ with its 
position vector $\boldsymbol{r}$ is in radial direction, and is a function
of $r$. Whereas the radiation pressure from the accretion disk is assumed to be a function
of $r$ and $\theta'$ where the latter is the angle between $\boldsymbol{J}_{\mathrm{D}}$
and $\boldsymbol{r}$ (see Secs.~\ref{sub:Continuum-Radiation-Source}
and \ref{sub:Precessing-Disk} for details). The point-source like
approximation for the disk radiation pressure at $P$ is valid when
$r_{\mathrm{D}}\ll r_{\mathrm{i}}$. Note that the figure is not to
scale. }

\label{fig:model-config}
\end{figure*}

Our basic model configuration is shown in Figure~\ref{fig:model-config}.
The model geometry and the assumptions of the SMBH and the disk are
very similar to those in \citet{Proga:2007}. In
Figure~\ref{fig:model-config}, a SMBH with its mass
$M_{\mathrm{BH}}$ and its Schwarzschild radius $r_{\mathrm{S}}=2GM_{\mathrm{BH}}/c^{2}$
is placed at the center of the cartesian coordinate system ($x$,
$y$, $z$). The X-ray emitting corona regions is defined as a sphere
with its radius $r_{*}$, as shown in the figure. The
geometrically-thin and optically-thick flat accretion disk (e.g.,
\citealt{shakura:1973})
is placed \emph{near} the $x$-$y$ plane.
In case for an axisymmetric model, the $z$-axis in the figure becomes
the symmetry axis, and the accretion disk is on the $x$-$y$ plane.
To simulate the disk precession, we assume that the angular momentum
($\boldsymbol{J}_{\mathrm{D}}$) of the accretion disk is tilted from
the $z$-axis by an angle $\Theta$. In other words,
the accretion disk is assumed to be tilted by $\Theta$ from the
$x$-$y$ plane. Further, the accretion disk hence its 
angular momentum $\boldsymbol{J}_{\mathrm{D}}$ is assumed to precess
around the $z$-axis with the precession period $P$. The 3-D hydrodynamic
simulations will be performed in the spherical coordinate system ($r$,
$\theta$, $\phi$), and in between the inner boundary $r_{\mathrm{i}}$
and the outer boundary $r_{\mathrm{o}}$. The poles of the spherical
coordinate system coincides with the $z$-axis. 
The radiation forces, from
the corona region (the sphere with its radius $r_{*}$) and the accretion
disk, acting on the gas located at a location ($p$) in the field
are assumed to be only in radial direction. The magnitude of the radiation
force due to the corona is assumed to be a function of radius $r$,
but that due to the accretion disk is assumed to be a function of
$r$ and the angle ($\theta'$) between the disk angular momentum
$\boldsymbol{J}_{\mathrm{D}}$ and the position vector $\boldsymbol{r}$
as shown in the figure. The point-source like approximation for the
disk radiation pressure at $P$ is valid when $r_{\mathrm{D}}\ll r_{\mathrm{i}}$.
In the following, we will describe our radiation hydrodynamics, our
implementation of the continuum radiation sources (the corona and
disk), the model parameters and assumptions. 


%


\subsection{Hydrodynamics}

\label{sub:Hydrodynamics}

We employ 3-D hydrodynamical simulations of the outflow from and accretion
onto a central part of AGN, using the ZEUS-MP code \citep[c.f.,][]{Hayes:2006}
which is a massively parallel MPI-implemented version of the ZEUS-3D
code (c.f., \citealt{Hardee:1992}; \citealt{Clarke:1996}). The ZEUS-MP
is a Eulerian hydrodynamics code which uses the method of finite differencing
on a staggered mesh with a second-order-accurate, monotonic advection
scheme \citep{Hayes:2006}. To compute the structure and evolution
of a flow irradiated by a strong continuum radiation of AGN, we solve
the following set of HD equations: 
\begin{eqnarray}
  \frac{D\rho}{Dt}+\rho\,\boldsymbol{\nabla}\cdot\boldsymbol{v} 
  & = & 0, 
  \label{eq:hydro01}
\end{eqnarray}

\begin{equation}
  \rho\frac{D\boldsymbol{v}}{Dt}=-\boldsymbol{\nabla}P+\rho\,\boldsymbol{g}+\rho\,\boldsymbol{g}_{\mathrm{rad}},
\label{eq:hydro02}
\end{equation}

\begin{equation}
  \rho\frac{D}{Dt}\left(\frac{u}{\rho}\right)=-P\,\boldsymbol{\nabla}\cdot\boldsymbol{v}+\rho\,\mathcal{C},
  \label{eq:hydro03}
\end{equation}
 where $\rho$, $u$, $P$ and $\boldsymbol{v}$ are the mass density,
energy density, pressure, and the velocity of gas respectively. Also,
$\boldsymbol{g}$ is the gravitational force per unit mass. The Lagrangian/co-moving
derivative is defined as $D/Dt\equiv\partial/\partial t+\boldsymbol{v}\cdot\boldsymbol{\nabla}$.
We have introduced two new components to the ZEUS-MP in order to treat
the gas dynamics more appropriate for the gas flow in and around AGN.
The first is the acceleration due to radiative force per unit mass
($\boldsymbol{g}_{\mathrm{rad}}$) in equation~(\ref{eq:hydro02}),
and the second is the the effect of radiative cooling (and heating)
simply as the net cooling rate ($\mathcal{C}$) in equation~(\ref{eq:hydro03}).
As this is our first 3-D simulations with this code, we consider a
simplest case i.e., $\mathcal{C}=0$ , but $g_{\mathrm{rad}}\neq0$.
We also use $\gamma=1.01$ in the equation of state $P=\left(\gamma-1\right)u$
where $\gamma$ is the adiabatic index. In the following, our implementation
of $\boldsymbol{g}_{\mathrm{rad}}$ will be described.

\subsubsection{Radiation Force}

To evaluate the radiative acceleration due to line absorption/scattering,
we follow the method in \citet{Proga:2000} who applied the modified
Castor, Abbott \& Klein (CAK) approximation \citep{Castor:1975}.
Their model works under the assumption of the Sobolev approximation
(e.g., \citealt{Sobolev:1957}; \citealt{Castor:1970}; \citealt{Lucy:1971});
hence, the following conditions are assumed to be valid: (1)~presence
of large velocity gradient in the gas flow, and (2)~the intrinsic
line width is negligible compared to the Doppler broadening of a line.
Following \citet{Proga:2000}, the radiative acceleration of a unit
mass at a point $\boldsymbol{r}$ can be written as 
\begin{equation}
  \boldsymbol{g}_{\mathrm{rad}}=\oint_{\Omega}\left[1+\mathcal{M}\right]\left[\frac{\sigma_{e}I\left(\boldsymbol{r},\hat{\boldsymbol{n}}\right)}{c}\right]\hat{\boldsymbol{n}}\,
 d\Omega 
\label{eq:rad_force}
\end{equation}
 where $I$ is the frequency-integrated continuum intensity in the
direction $\hat{\boldsymbol{n}}$, and $\Omega$ is the solid angle
subtended by the source of continuum radiation. Also, $\sigma_{e}$
is the electron scattering cross section. The force multiplier $\mathcal{M}$
is a function of optical depth parameter $\tau$ which is similar
to the Sobolev optical depth (c.f.~\citealt{Rybicki:1978}), and
can be written as \begin{equation}
\tau=\frac{\sigma_{e}\,\rho\, v_{\mathrm{th}}}{\left|Q\right|}\label{eq:tau_parameter}\end{equation}
where $Q=dv_{l}/dl$ is the directional derivative of the velocity
field in direction $\hat{\boldsymbol{n}}$, $dl$ is the line element
in the same direction, and $v_{\mathrm{th}}$ is the thermal velocity
of the gas. Further equation~(\ref{eq:rad_force}) can be simplified
greatly when the continuum radiation source is approximated as a point,
i.e.,~when $r\gg r_{\mathrm{c}}$ where $r_{\mathrm{c}}$ is the
radius of the radiation source. In our case, we consider the accretion
disk which emits most of the radiation from the innermost part, between
$r_{*}$ and $r_{\mathrm{D}}$ in Figure~\ref{fig:model-config};
hence, the condition $r\gg r_{\mathrm{c}}$ is satisfied. Using this
approximation, the radiative acceleration $\boldsymbol{g}_{\mathrm{rad}}$
will be radial only, and be a function of radial position and polar
angle (if the contribution from the disk luminosity is included),
i.e. $\boldsymbol{g}_{\mathrm{rad}}=g_{\mathrm{rad}}\left(r,\theta\right)\hat{\boldsymbol{r}}$.
This simplification is very useful for our purposes as it reduces
the computational time significantly hence it enables us to perform
large-scale 3-D simulations. Unlike \citet{Proga:2000}, we consider
the case in which the radiative acceleration is dominated by the continuum
process, i.e. $\mathcal{M}=0$ in equation~(\ref{eq:rad_force})
in this paper since we initially intend to investigate the basic characteristics
of the impact of the disk precession that do not depend on the details
of the radiation force model. The models with $\mathcal{M}\neq0$
in equation~(\ref{eq:rad_force}) and $\mathcal{C}\neq0$ will be
presented in a forthcoming paper.

\subsection{Continuum Radiation Source }

\label{sub:Continuum-Radiation-Source}

As mentioned earlier, we consider two different continuum radiation
sources in our models: (1)~the accretion disk, and (2)~the central
spherical corona. Since the geometry of the central engine in AGN
is not well understood, we assume that is consist of a spherically
shaped corona with its radius $r_{*}$ and the innermost part of the
accretion disk (c.f.~Fig.~\ref{fig:model-config}). The disk is
assumed to be flat, Keplerian, geometrically-thin and optically thick.
The disk radiation is assumed to be dominated by the radiation from
the disk radius between $r_{*}$ and $r_{\mathrm{D}}$ where $r_{*}=3\, r_{\mathrm{S}}$
and $r_{*}<r_{\mathrm{D}}\ll r_{\mathrm{i}}$ (c.f., Fig.~\ref{fig:model-config}).
Note that the exact size of $r_{\mathrm{D}}$ does not matter as
long as it satisfies this condition in order for the point-source
approximation mentioned in \S.~\ref{sub:Hydrodynamics} to be valid. 

In terms of the disk mass-accretion rate ($\dot{M}_{\mathrm{D}}$),
the mass of the BH ($M_{\mathrm{BH}}$) and the Schwarzschild radius
($r_{\mathrm{S}}$), the total luminosity ($L$) of the system can
be written as 
\begin{align}
L & =\eta\dot{M}_{\mathrm{D}}c^{2}\label{eq:total-luminosity}\\
  & =\frac{2\eta GM_{\mathrm{BH}}\dot{M}_{\mathrm{D}}}{r_{\mathrm{S}}}\label{eq:total-luminosity2}
\end{align}
 where $\eta$ is the rest mass conversion efficiency (e.g., \citealt{shakura:1973}).
Following \citet{Proga:2007} and \citet{Proga:2007b}, we simply
assume the system essentially radiates only in the UV and the X-ray
bands. The total luminosity of the system $L$ is then the sum of
the UV luminosity $L_{\mathrm{UV}}$ and the X-ray luminosity $L_{\mathrm{X}}$
i.e., $L=L_{\mathrm{UV}}+L_{\mathrm{X}}$. Further, we assume that
the disk only radiates in the UV and the central corona in the X-ray.
The ratio of the disk luminosity ($L_{\mathrm{D}}$) to the total
luminosity is parametrized as $f_{\mathrm{D}}=L_{\mathrm{D}}/L$,
and that of the corona luminosity ($L_{*}$) to the total luminosity
as $f_{*}=L_{*}/L$. Consequently, $f_{\mathrm{D}}+f_{*}=1$.

In the point-source approximation limit, the radiation flux from the
central X-ray corona region can be written as 
\begin{equation}
  \mathcal{F}_{*}=\frac{L_{*}}{4\pi r^{2}}
  \label{eq:corona-flux}
\end{equation}
 where $r$ is the radial distance from the center (by neglecting
the source size). Here we neglect the geometrical obscuration of the
corona emission by the accretion disk and vice versa. On the other
hand, the disk radiation depends on the polar angle $\theta$ because
of the source geometry. Again following \citet{Proga:2007} and
\citet{Proga:2007b} (see also \citealt{proga:1998}), the disk intensity
$I_{\mathrm{D}}$ is assumed to be radial and $I_{\mathrm{D}}\propto\left|\cos\theta'\right|$.
This follows that the disk radiation flux at the distance $r$ from
the center can be written as 
\begin{equation}
\mathcal{F}_{\mathrm{D}}=2\,\left|\cos\theta'\right|\,\frac{L_{\mathrm{D}}}{4\pi
  r^{2}}
\label{eq:disk-flux}
\end{equation}
 where $\theta'$ is the angle between the disk normal and the position
vector $\boldsymbol{r}$ (c.f., Fig.~\ref{fig:model-config}). The leading term $2$ in this expression
comes from the normalization of the polar angle dependency. Finally
by using eqs.~(\ref{eq:rad_force}), (\ref{eq:corona-flux}) and (\ref{eq:disk-flux}),
the radiative acceleration term in equation~(\ref{eq:hydro03}) can
be written as 
\begin{equation}
   \boldsymbol{g}_{\mathrm{rad}}=\frac{\sigma_{e}L}{4\pi r^{2}c}\left\{
f_{*}+2\,\left|\cos\theta'\right|\, f_{\mathrm{D}}\right\}
\,\boldsymbol{\hat{r}}\, .
\label{eq:rad-force-final}
\end{equation}

\subsection{Precessing Disk}

\label{sub:Precessing-Disk}

As we noted before, here we do not model the precession of the accretion
disk itself, but rather manually force the precession. We do not specify
the cause of the precession either. We simply assume that the disk
precession exists, and investigate its consequence to the AGN environment.
The UV emitting portion of the disk spans from $r_{*}$ to $r_{\mathrm{D}}$
(c.f., Fig.~\ref{fig:model-config}). We assume that $r_{\mathrm{D}}\ll r_{\mathrm{i}}$
where the $r_{\mathrm{i}}$ is the inner radius of the computational
domain of the hydrodynamic simulations. This means that the disk itself
is not in the computational domain. The effect of the precessing disk
is included as precessing radiation field in the hydrodynamics of
the gas (through eq.~{[}\ref{eq:hydro02}]). 

We assume that the disk is tilted from the $x$-$y$ plane (in the
cartesian coordinate system) by $\Theta$ as in Figure~\ref{fig:model-config}.
Equivalently, the disk angular momentum $\boldsymbol{J}_{\mathrm{D}}$
(assuming a flat uniform Keplerian disk) deviates from the $z$-axis
by $\Theta$. Further, we assume that $\boldsymbol{J}_{\mathrm{D}}$
precesses around the $z$-axis with precession period $P$. With these
assumptions, the components of the $\boldsymbol{J}_{\mathrm{D}}$
in the cartesian coordinate system can be written as 
\begin{align}
  J_{\mathrm{D}x} & =J_{D}\sin\Theta\,\cos\left(\frac{2\pi t}{P}\right),\label{eq:J_D-x-component}\\
  J_{\mathrm{D}y} & =J_{D}\sin\Theta\,\sin\left(\frac{2\pi t}{P}\right),\label{eq:J_D-y-component}\\
  J_{\mathrm{D}z} & =J_{D}\cos\Theta\label{eq:J_D-z-component}
\end{align}
where $t$ is the time measured from the beginning of hydrodynamic
simulations. Here we set $\boldsymbol{J}_{\mathrm{D}}$ to be on the
$x-z$ plane (as shown in Fig.~\ref{fig:model-config}) at $t=0$.
By setting $\Theta=0$, the model reduced to an asymmetric case as
in \citet{Proga:2007}. To compute the radiative acceleration as expressed
in equation~(\ref{eq:rad-force-final}), one requires the angle between
$\boldsymbol{J}_{\mathrm{D}}$ and the position vector $\boldsymbol{r}$
at which the set of the HD equations (eqs.~{[}\ref{eq:hydro01}],
{[}\ref{eq:hydro02}] and {[}\ref{eq:hydro03}]) are solve. This can
be obtained simply by finding the inner product of $\boldsymbol{J}_{\mathrm{D}}$
and $\boldsymbol{r}$.

\subsection{Model Setup}

\label{sub:Model-Setup}

In all models presented here, the following ranges of the coordinates
are adopted: $r_{\mathrm{i}}\leq r\leq r_{\mathrm{o}}$, $0\leq\theta\leq\pi$
and $0\leq\phi<2\pi$ where $r_{\mathrm{i}}=500\, r_{*}$ and $r_{\mathrm{o}}=2.5\times10^{5}\, r_{*}$.
The radius of the central and spherical X-ray corona region $r_{*}$
coincides with the inner radius of the the accretion disk (Fig.~\ref{fig:model-config}).
In our simulations, the polar and azimuthal angle ranges are divided
into 128 and 64 zones, and are equally spaced. In the $r$ direction,
the gird is divided into 128 zones in which the zone size ratio is
fixed at $\Delta r_{k+1}/\Delta r_{k}=1.04$. 

For the initial conditions, the density and the temperature of gas
are set uniformly i.e., $\rho=\rho_{o}$ and $T=T_{o}$ everywhere
in the computational domain where $\rho_{o}=1.0\times10^{-21}\,\mathrm{g\, cm^{-3}}$
and $T_{o}=2\times10^{7}\,\Kelvin$ through out this paper. The initial
velocity of the gas is simply set to zero everywhere. 

At the inner and outer boundaries, we apply the outflow (free-to-outflow)
boundary conditions, in which the field values are extrapolated beyond
the boundaries using the values of \emph{the ghost zones} residing
outside of normal computational zones (see \citealt{Stone:1992} for
more details). At the outer boundary, all HD quantities (except the
radial velocity) are fixed constant, to their initial values (e.g., $T=T_{o}$
and $\rho=\rho_{o}$), during the the evolution of each model. The
radial velocity components are allowed to float. \citet{Proga:2007} applied
these conditions to represent a steady flow condition at the outer
boundary. They found that this technique leads to a solution that
relaxes to a steady state in both spherical and non-spherical accretion
with an outflow (see also \citealt{Proga:2003b}). This imitates the
condition in which a continuous supply of gas is available at the
outer boundary.

\section{Results}

\label{sec:Results}

\subsection{Reference Values}

\label{sub:Reference-Values}


%
\begin{table*}

\caption{\label{tab:Model-Summary}Model Summary}

\scriptsize

\begin{center}

\begin{tabular}{cccccccccc}
\hline 
\hline&
$\Theta$&
$P$&
$\dot{M}_{\mathrm{in}}\left(r_{\mathrm{o}}\right)$&
$\dot{M}_{\mathrm{net}}\left(r_{\mathrm{i}}\right)$&
$\dot{M}_{\mathrm{out}}\left(r_{\mathrm{o}}\right)$&
$v_{r}^{\mathrm{max}}\left(r_{\mathrm{o}}\right)$&
$P_{k}\left(r_{\mathrm{o}}\right)$&
$P_{\mathrm{th}}\left(r_{o}\right)$&
$j_{\rho}$\tabularnewline
Model&
$\left(\,^{\circ}\right)$&
$\left(\mathrm{yr}\right)$&
$\left(10^{25}\mathrm{\, g\, s^{-1}}\right)$&
$\left(10^{25}\,\mathrm{g\, s^{-1}}\right)$&
$\left(10^{25}\,\mathrm{g\, s^{-1}}\right)$&
$\left(\mathrm{km\, s^{-1}}\right)$&
$\left(10^{40}\,\mathrm{erg\, s^{-1}}\right)$&
$\left(10^{40}\,\mathrm{erg\, s^{-1}}\right)$&
$\left(j_{0}\right)$\tabularnewline
\hline
I&
$0$&
$\infty$&
$-2.2$&
$-0.6$&
$1.6$&
$1500$&
$3.0$&
$260$&
$0$\tabularnewline
II&
$5$&
$1.6\times10^{4}$&
$-2.3$&
$-0.6$&
$1.7$&
$640$&
$1.4$&
$290$&
$0.2$\tabularnewline
III&
$15$&
$1.6\times10^{4}$&
$-2.2$&
$-0.4$&
$1.8$&
$620$&
$1.1$&
$310$&
$0.2$\tabularnewline
IV&
$5$&
$1.6\times10^{5}$&
$-2.2$&
$-0.6$&
$1.6$&
$1900$&
$3.0$&
$260$&
$0.05$\tabularnewline
\hline
\end{tabular}

\end{center}

\tablecomments{The model output values are averaged over the last $2\times10^{12}\,\mathrm{s}$
of the hydrodynamic simulations. }

\normalsize
\end{table*}


We consider four different cases which have different combinations
of the disk tilt angle ($\Theta$) and the disk precession period ($P$),
as summarized in Table~\ref{tab:Model-Summary}. The following parameters
are common to all the models presented here, and are exactly the same
as in \citet{Proga:2007}. We assume that the central BH is non-rotating
and has mass $M_{\mathrm{BH}}=10^{8}\, M_{\odot}$. The size of the
disk inner radius is assumed to be $r_{*}=3r_{s}=8.8\times10^{13}\,\mathrm{cm}$
(c.f.~Sec.~\ref{sub:Model-Setup}). The mass accretion rate ($\dot{M}_{a}$)
onto the central SMBH and the rest mass conversion efficiency ($\eta$)
are assumed to be $1\times10^{26}\,\mathrm{g\, s^{-1}}$ and $0.0833$,
respectively. With these parameters, the corresponding accretion luminosity
of the system is $L=7.5\times10^{45}\,\mathrm{erg\, s^{-1}=2\times10^{12}\,\Lsun}$.
Equivalently, the system has the Eddington number $\Gamma=0.6$ where
$\Gamma\equiv L/L_{\mathrm{Edd}}$ and $L_{\mathrm{Edd}}$ is the
Eddington luminosity of the Schwarzschild BH i.e., $4\pi cGM_{\mathrm{BH}}/\sigma_{e}$.
The fractions of the luminosity in the UV ($f_{\mathrm{UV}}$) and
that in the X-ray ($f_{\mathrm{X}}$) are fixed at $0.95$ and $0.05$
respectively, as in \citet{Proga:2007} (see their Run~C). 

Important reference physical quantities relevant to our systems are
as follows. The Compton radius, $R_{C}\equiv GM_{\mathrm{BH}}\mu\, m_{p}/kT_{C}$,
is $8\times10^{18}\,\mathrm{cm}$ or equivalently $9\times10^{4}\, r_{*}$
where $T_{C}$, $\mu$ and $m_{p}$ are the Compton temperature, the
mean molecular weight of gas and the proton mass, respectively. Here
we assume that the gas temperature at infinity is $T_{\infty}=T_{C}=2\times10^{7}\,\Kelvin$;
hence, the corresponding speed of sound at infinity is $c_{\infty}=(\gamma kT_{C}/\mu m_{p})^{1/2}=4\times10^{7}\,\mathrm{cm\, s^{-1}}$
where $\gamma$ is the adiabatic index. In this paper, $\gamma=1.01$
(almost isothermal) is adopted to imitate a gas in Compton equilibrium
with the radiation field. The corresponding Bondi radius \citep{Bondi:1952}
is $R_{B}=GM_{\mathrm{BH}}/c_{\infty}^{2}=4.8\times10^{18}\,\mathrm{cm}$
while its relation to the Compton radius is $R_{B}=\gamma^{-1}R_{C}$.
The Bondi accretion rate (for the isothermal flow) is $\dot{M}_{B}=3.3\times10^{25}\,\mathrm{g\, s^{-1}}=0.52\,\MsunPerYear$.
The corresponding free-fall time ($t_{\mathrm{ff}}$) of gas from
the Bondi radius to the inner boundary is $2.1\times10^{11}\,\mathrm{sec}=7.0\times10^{3}\,\mathrm{yr}$
which is about $2.3$ times smaller than the precession period used
for Models~II and III, and about $23$ times smaller than that of
Model~IV (c.f.~Tab.~\ref{tab:Model-Summary}).

\subsection{Comparison of axisymmetric models in 2-D and 3-D}

\label{sub:Comparison-2D-3D}

Before we proceed to the main precession disk models, we briefly compare
our axisymmetric model (Model~I) with the axisymmetric models presented
earlier by \citet{Proga:2007} who used very similar model parameters
as in our Model~I. The main differences here are in the treatment
of the radiation force and that in the radiative heating/cooling.
As mentioned earlier, we set the force multiplier $\mathcal{M}=0$
(in eq.~{[}\ref{eq:rad_force}]) and the net cooling rate $\mathcal{C}=0$
(in eq.~{[}\ref{eq:hydro03}]) while \citet{Proga:2007} used non-zero
values of those two terms. In our Model~I, the adiabatic index is
set to $\gamma=1.01$ (essentially isothermal), but their models use
$\gamma=5/3$. However, \citet{Proga:2007} found that their Run~A
is nearly isothermal despite $\gamma=5/3$ was used (see their Fig.~1).
Another important difference is the numerical codes used. \citet{Proga:2007}
used the ZEUS-2D code \citep{Stone:1992}. 

\begin{figure*}
\begin{center}

\begin{tabular}{cc}
\includegraphics[clip,width=0.52\textwidth]{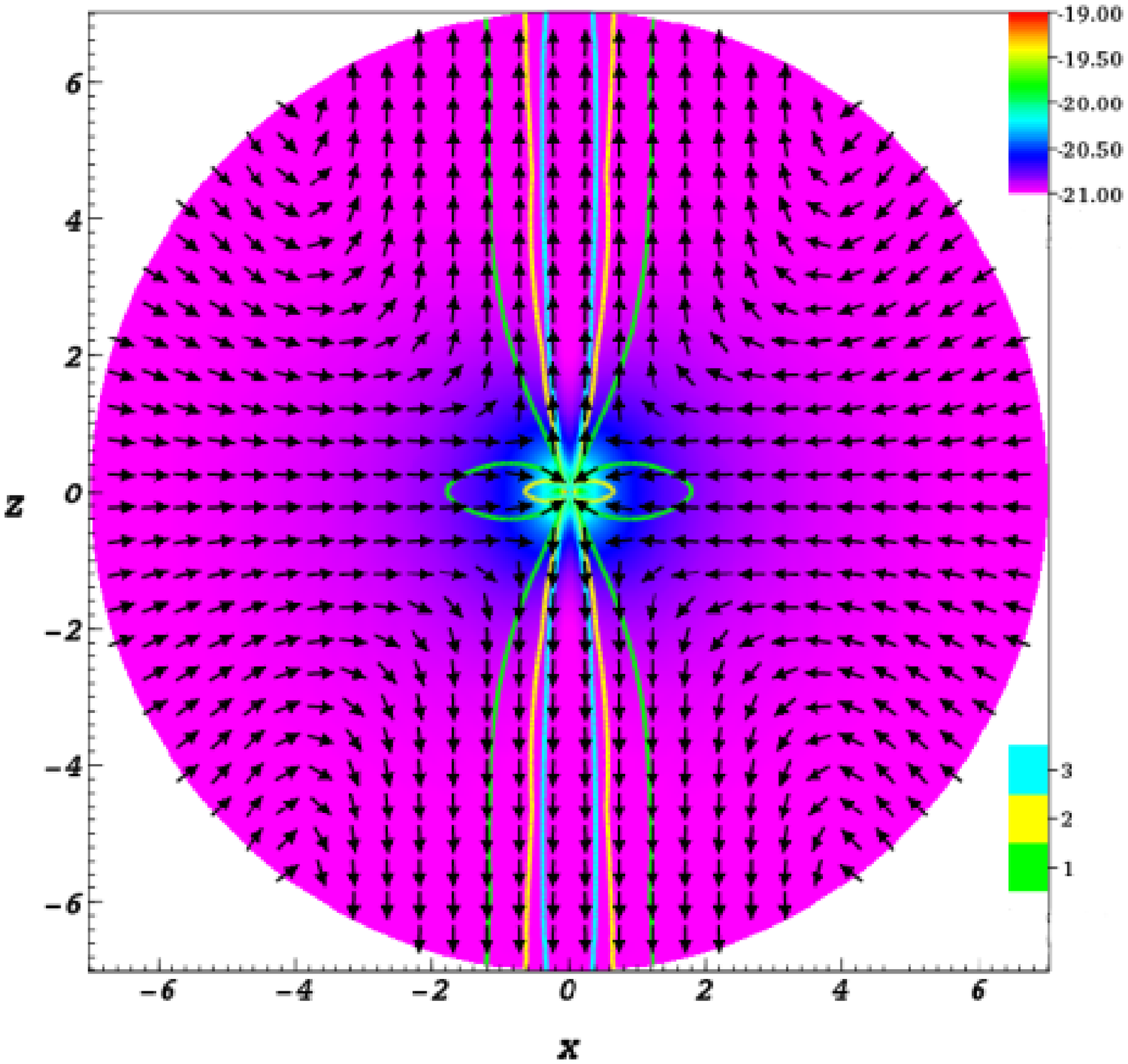}&
\hspace{-0.8 cm}\includegraphics[clip,width=0.52\textwidth]{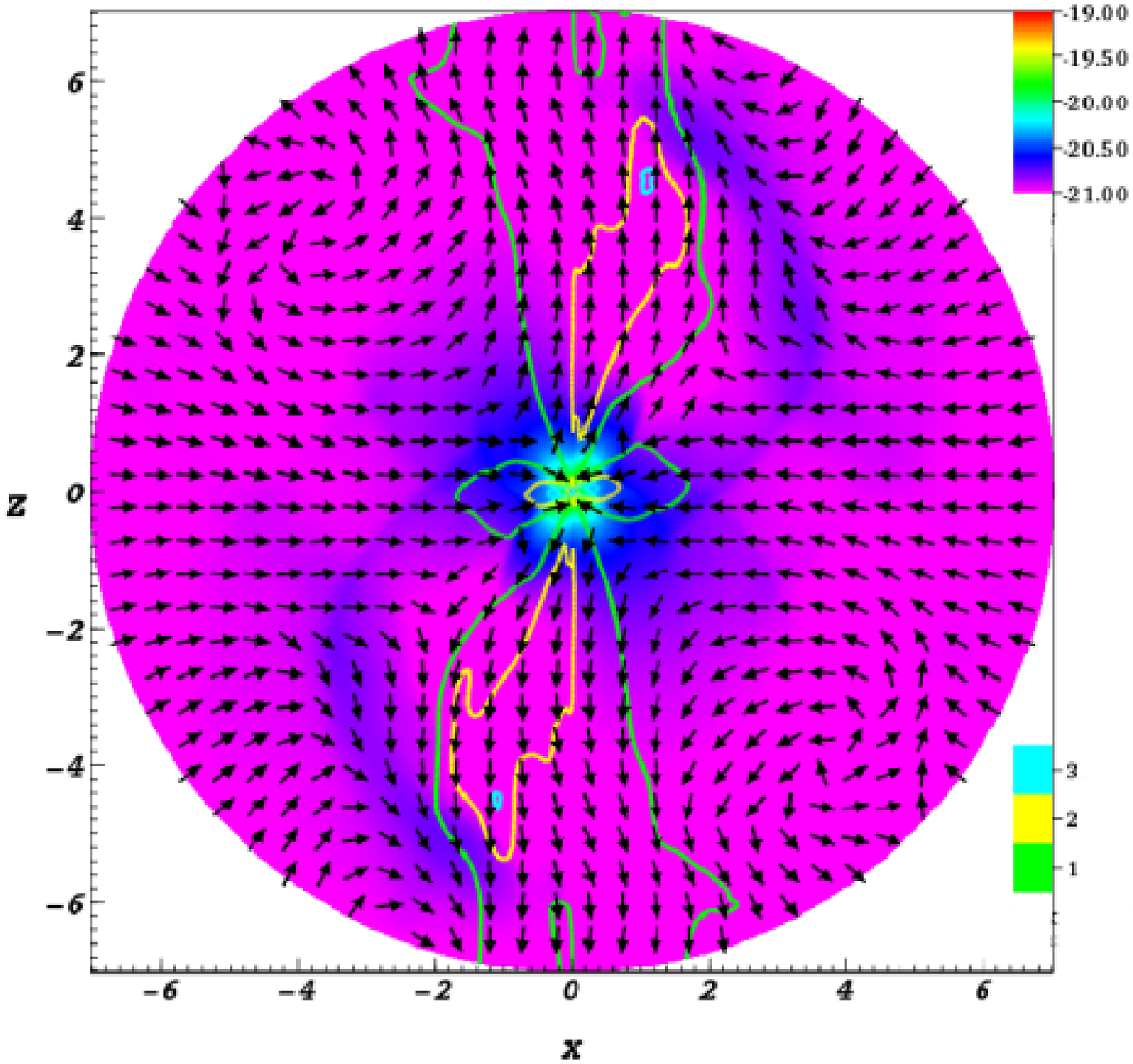}\tabularnewline
\includegraphics[clip,width=0.52\textwidth]{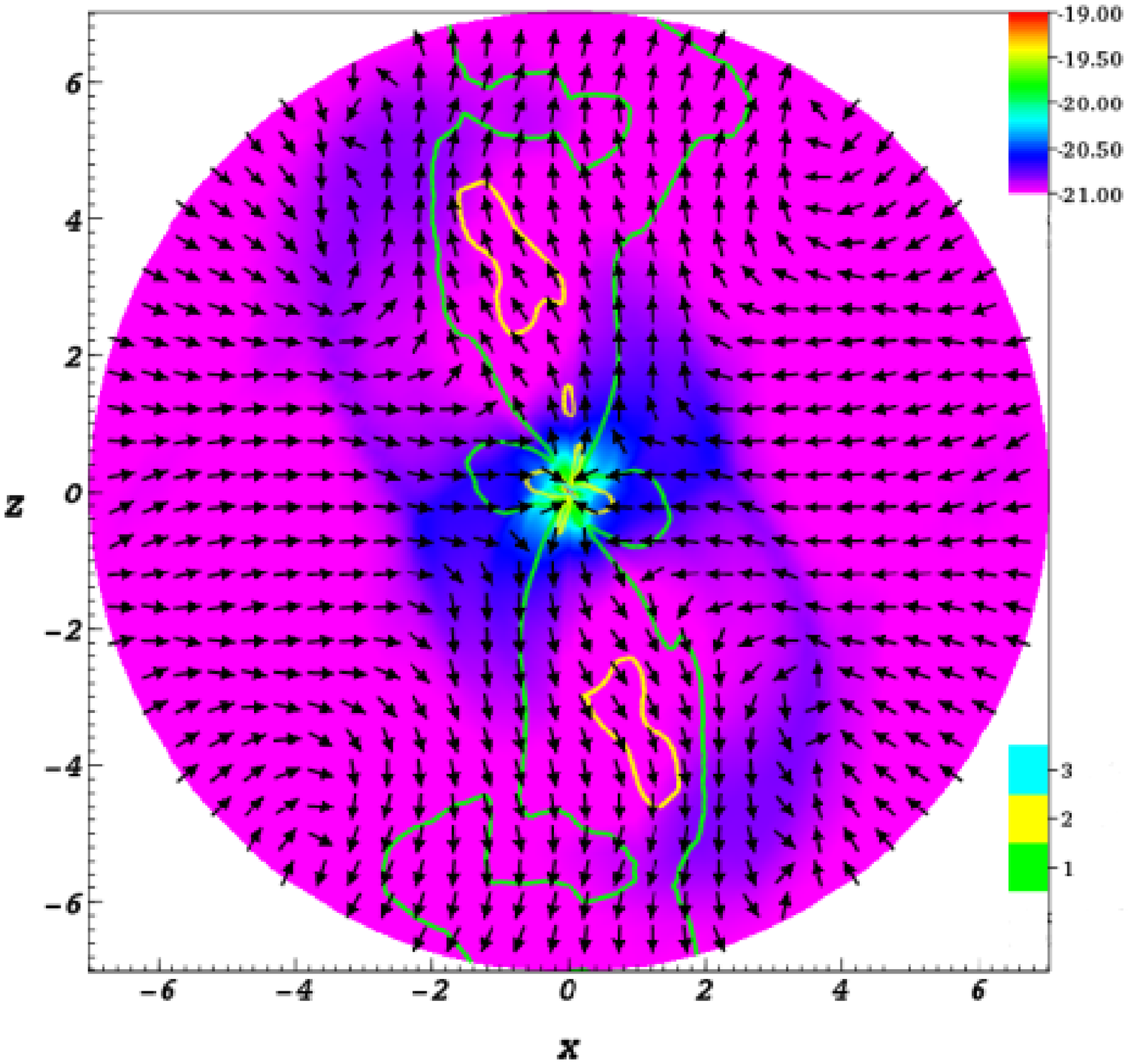}&
\hspace{-0.8 cm}\includegraphics[clip,width=0.52\textwidth]{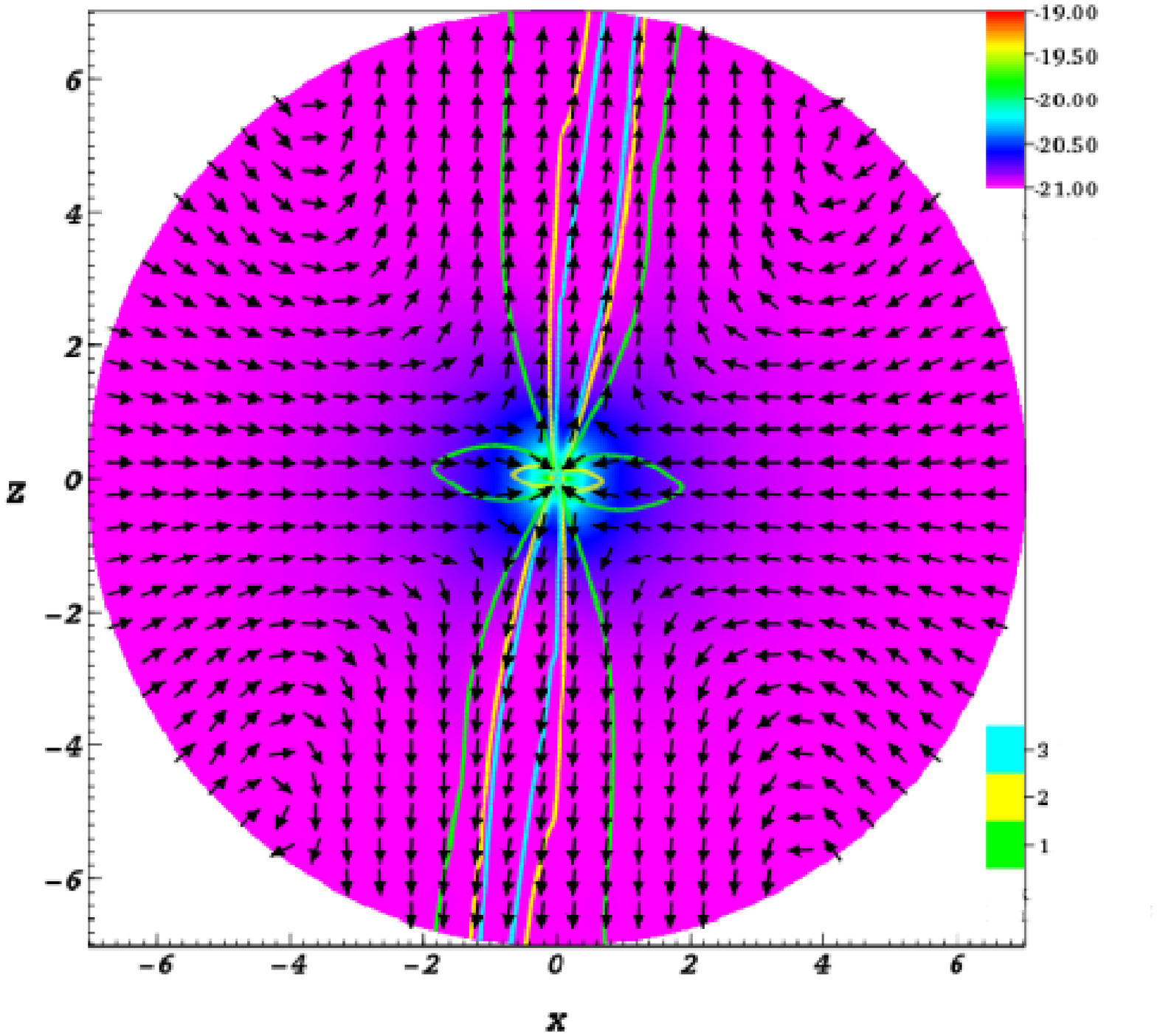}\tabularnewline
\end{tabular}

\end{center}

\caption{Comparison of the density and velocity fields on the $x$-$z$ plane
from Models~I (upper-left) II (upper-right), III (lower-left) and
IV (lower-right). The density maps shown in the background are given
in logarithmic scale (base 10) and in cgs units. The contours of the
Mach number are overlaid along with the arrows which indicate the
directions of the velocity on the x-$z$ plane. The units of both
$x$ and $z$ axes are in pc. The time slices of each models are chosen
such that the density and velocity fields are representative of each
model. }

\label{fig:Density-and-velocity}
\end{figure*}

\begin{figure*}
\begin{center}

\begin{tabular}{cc}
\includegraphics[clip,width=0.48\textwidth]{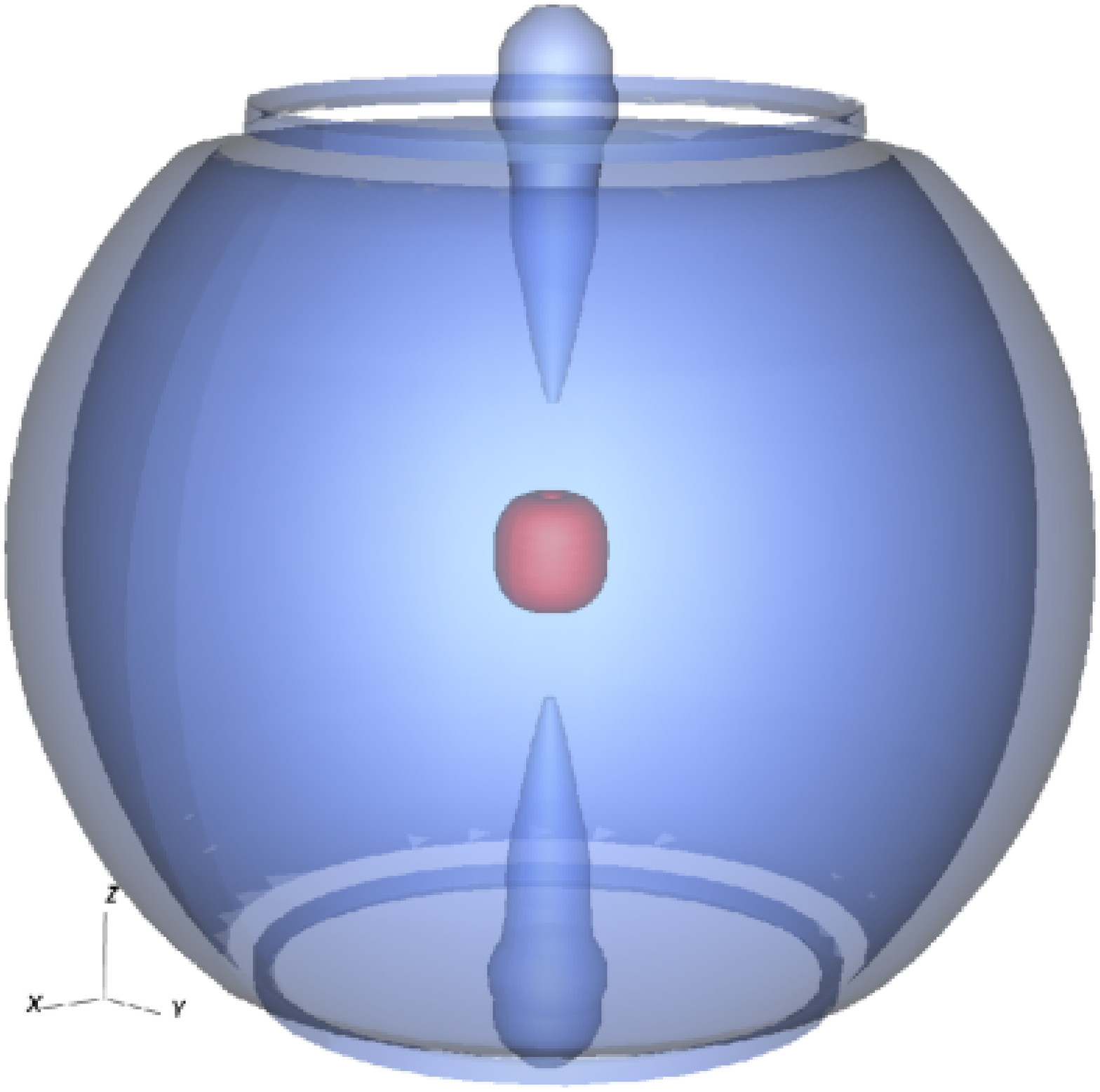}&
\includegraphics[clip,width=0.48\textwidth]{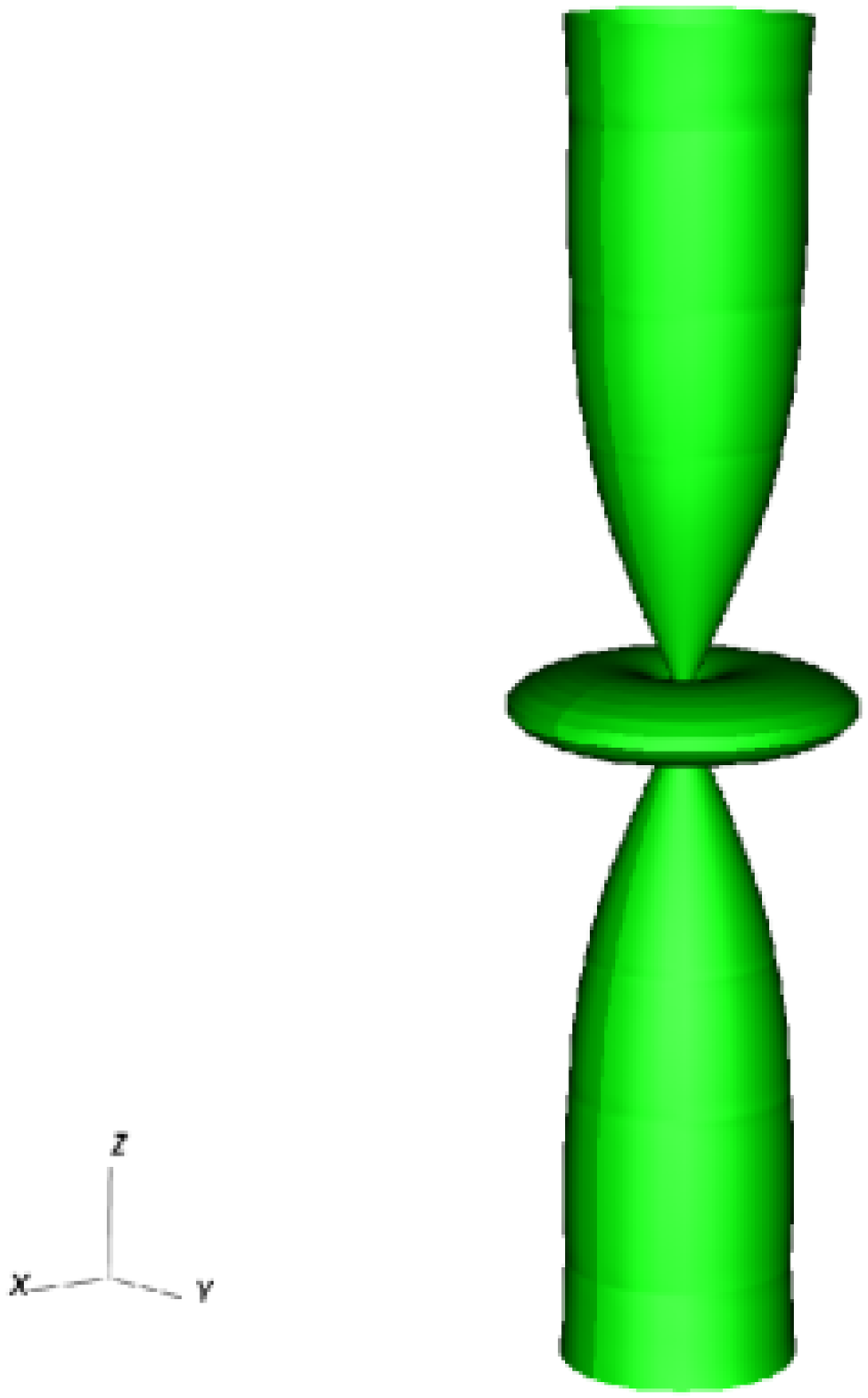}\tabularnewline
\includegraphics[clip,width=0.48\textwidth]{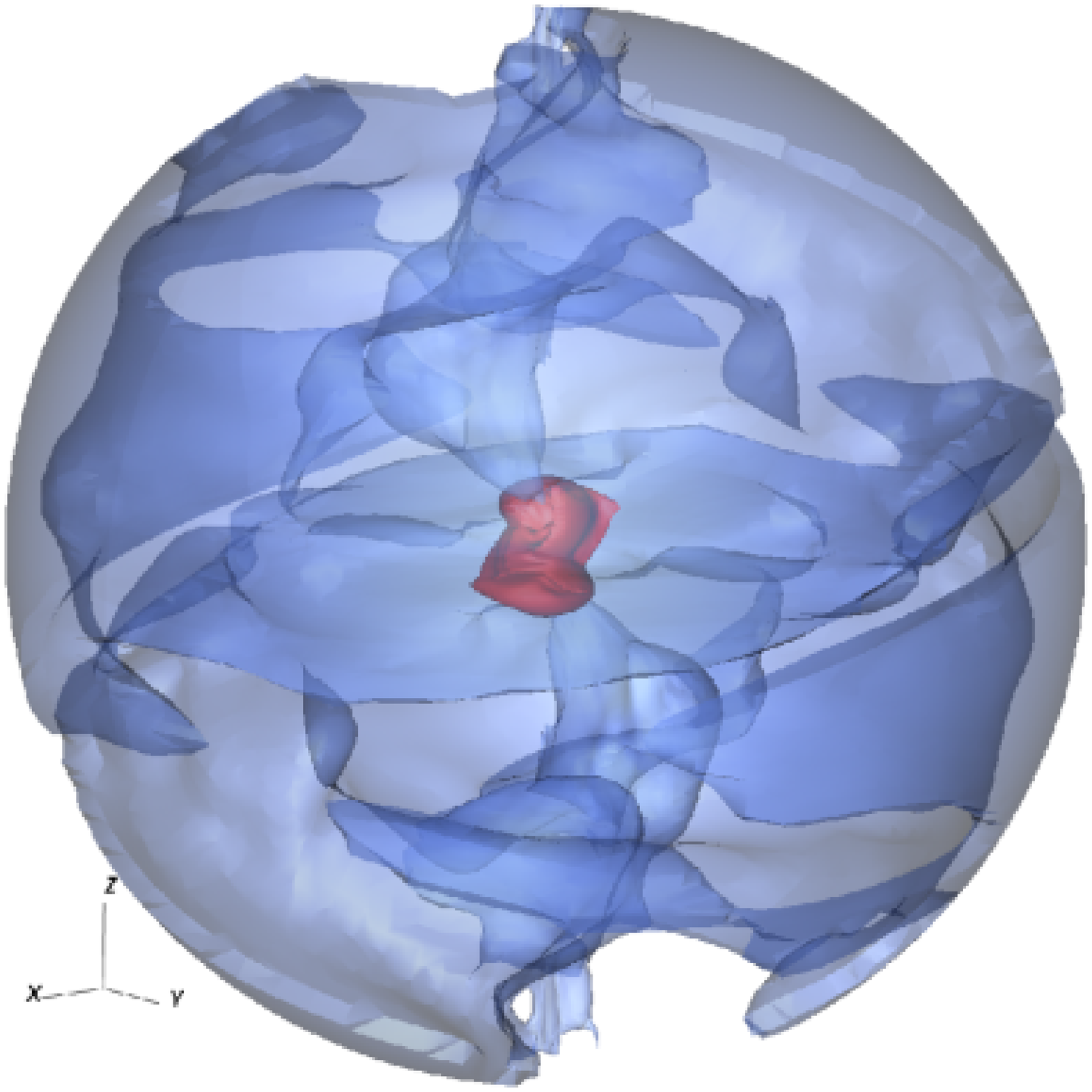}&
\includegraphics[clip,width=0.48\textwidth]{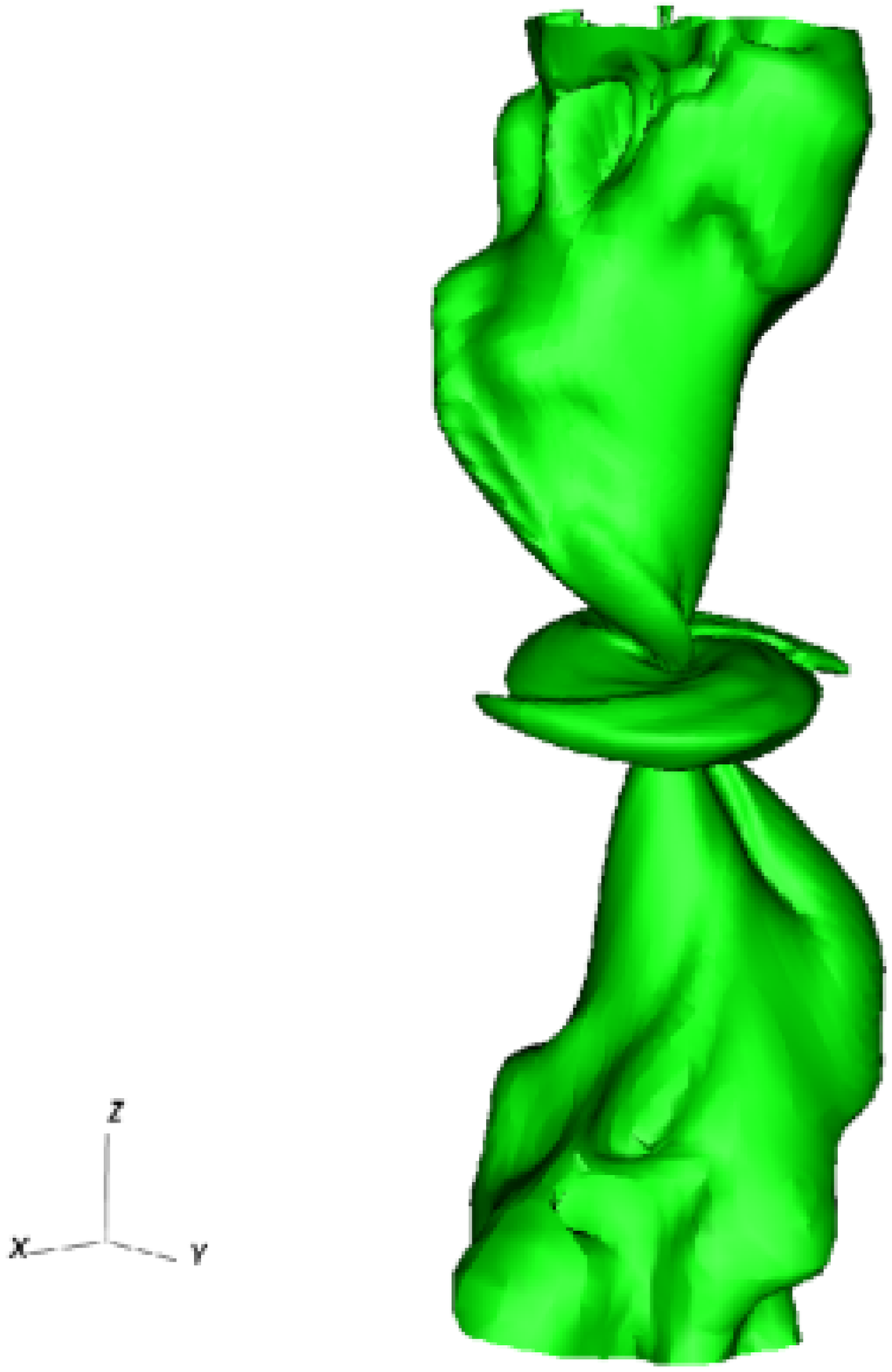}\tabularnewline
\end{tabular}

\end{center}

\caption{Two-level iso-density (left panels) and the corresponding sonic (right
panels) surfaces in 3-D for Models~I (upper panels) and II (lower
panels). The density levels used here are $\log\,\rho=-20.5$ (blue)
and $-21$ (red) where $\rho$ is in $\mathrm{g\, cm^{-3}}$. The
time slices of the simulation data used here are as in Fig.~\ref{fig:Density-and-velocity}.
The sizes of the plotting boxes are $14.2$~pc in all directions
($x$, $y$ and $z$). }

\label{fig:3-d-plots}
\end{figure*}

Overall geometry of the flow in Model~I (Figs.~\ref{fig:Density-and-velocity}
and \ref{fig:3-d-plots}) is similar to those in \citet{Proga:2007}.
The matter accretes onto the central BH near the equatorial plane,
and strong outflows occur in polar direction. The collimation of our
model is relatively weak compared to their Run~C which uses exactly
the same disk and corona luminosities as in our Model~I. The wider
bipolar outflow pattern seen here resembles that of their Run~A which
has the highest X-ray heating. The difference and the resemblance
seen here are caused by the following two key factors: (1) nearly
isothermal equation of state and (2) no radiative cooling ($\mathcal{C}=0$)
in our model . These condition keep the temperature of gas warm everywhere
in the computational domain, and the temperature is essentially that
set at the outer boundary ($T_{\infty}=T_{C}=2\times10^{7}\,\Kelvin$).
This will result in a very similar situation as in Run~A of \citet{Proga:2007}
in which the gas temperature is also relatively high because of the
high X-ray heating and cooling. The high temperature hence the ionization
state of the gas makes the line force in their model very inefficient,
resulting in the situation in which the gas is almost entirely driven
by the continuum process (electron scattering) and thermal effects
just as in our Model~I. 

Although not shown here, we have also checked the internal consistency
of the ZEUS-MP (3-D) code by running the axisymmetric models (Model~I)
in both 2-D and 3-D modes. We find that the results from the both
runs agree with each other in all aspects e.g., inflow and outflow
geometry, density distribution, velocity, mass accretion and outflow
rates. 


%



%



%
\begin{figure*}
\begin{center}

\begin{tabular}{cc}
\includegraphics[clip,width=0.48\textwidth]{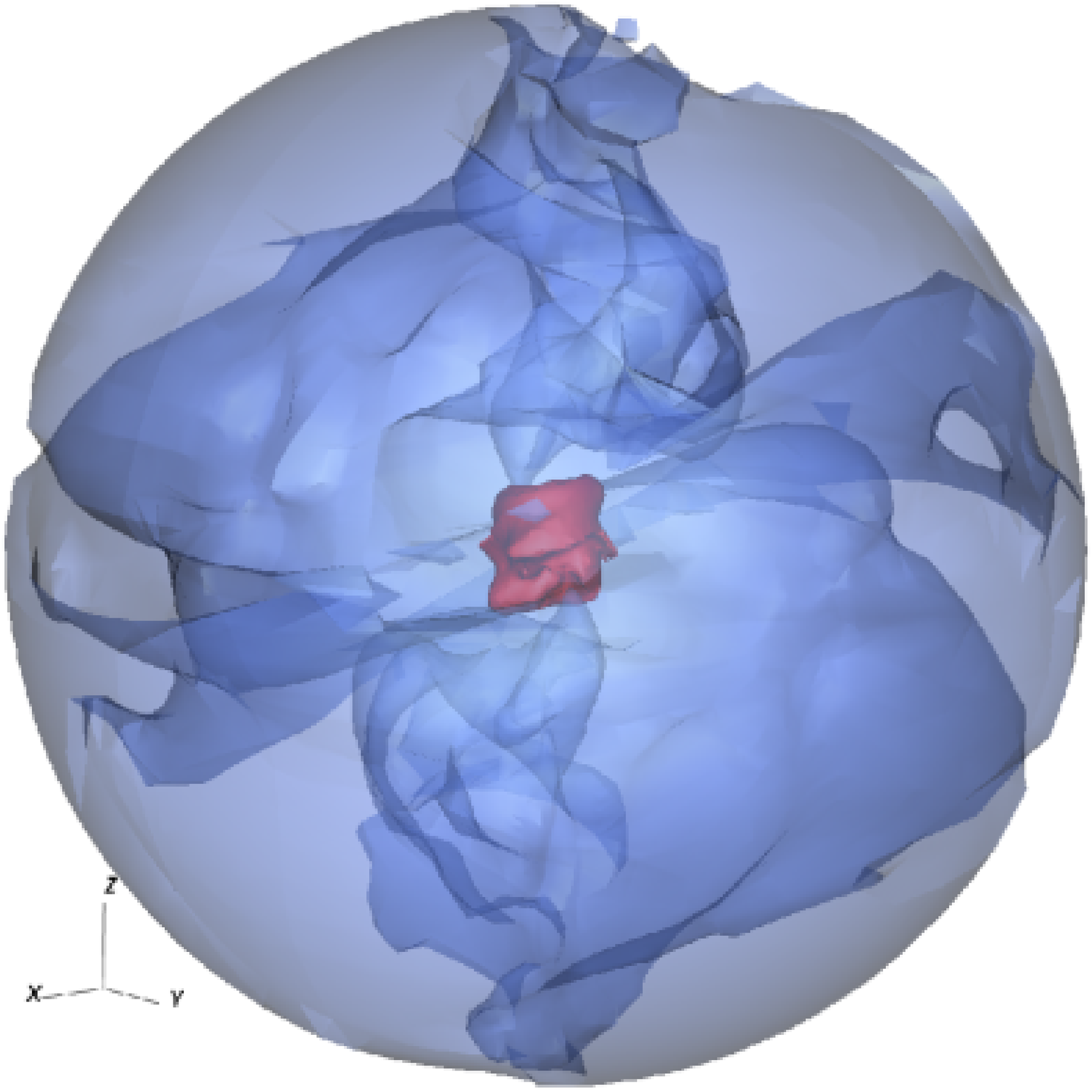}&
\includegraphics[clip,width=0.48\textwidth]{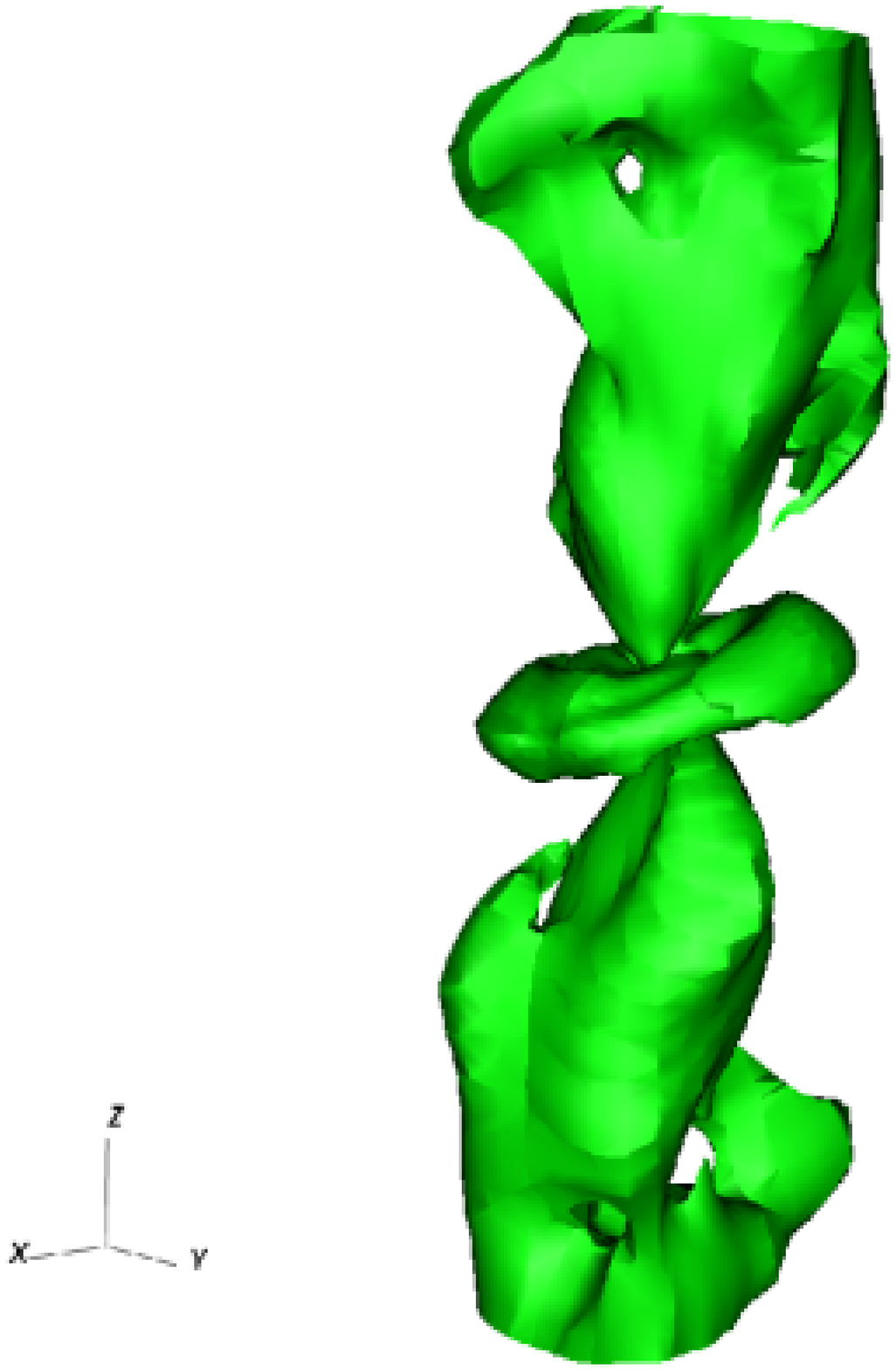}\tabularnewline
\includegraphics[clip,width=0.48\textwidth]{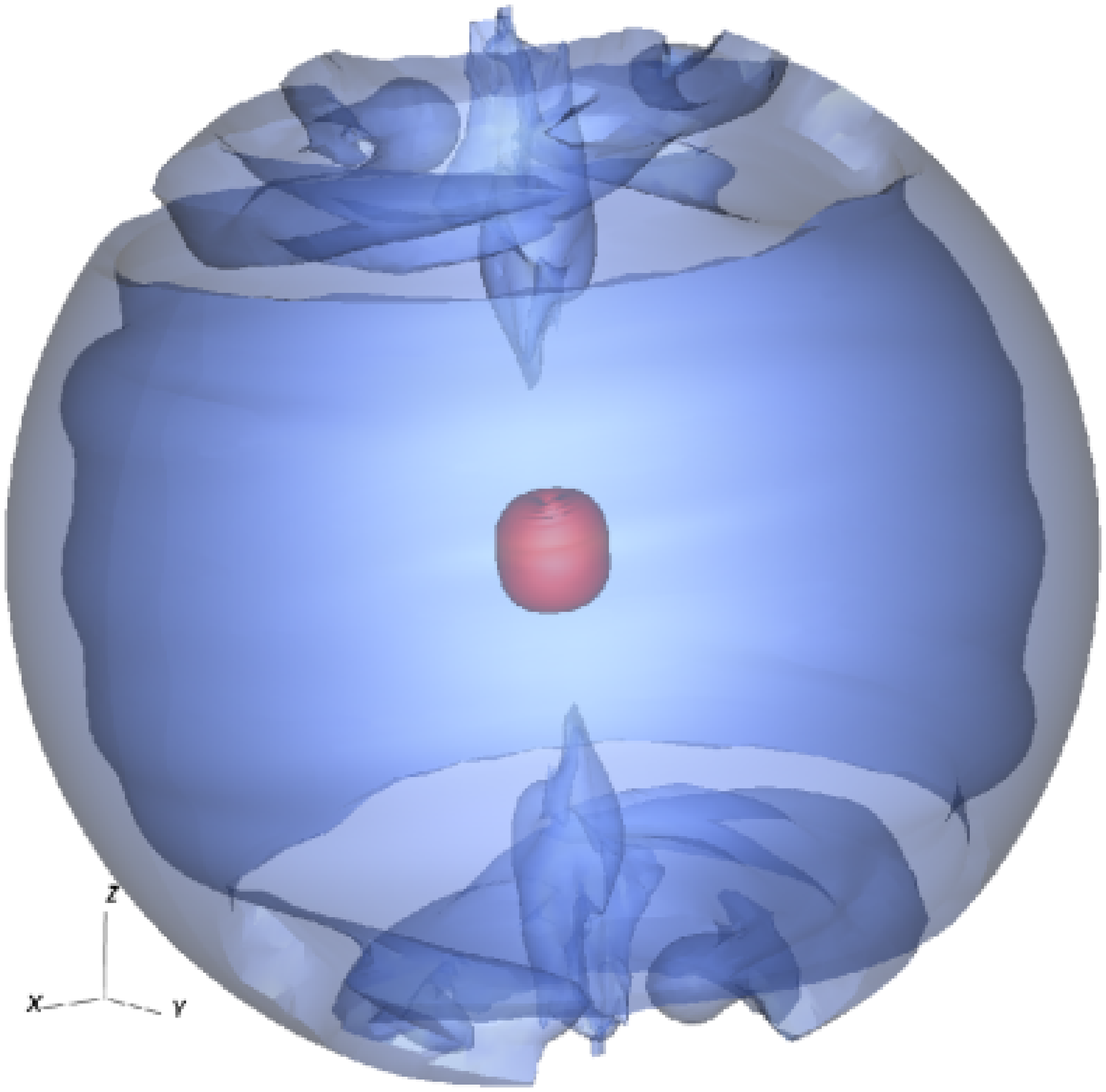}&
\includegraphics[clip,width=0.48\textwidth]{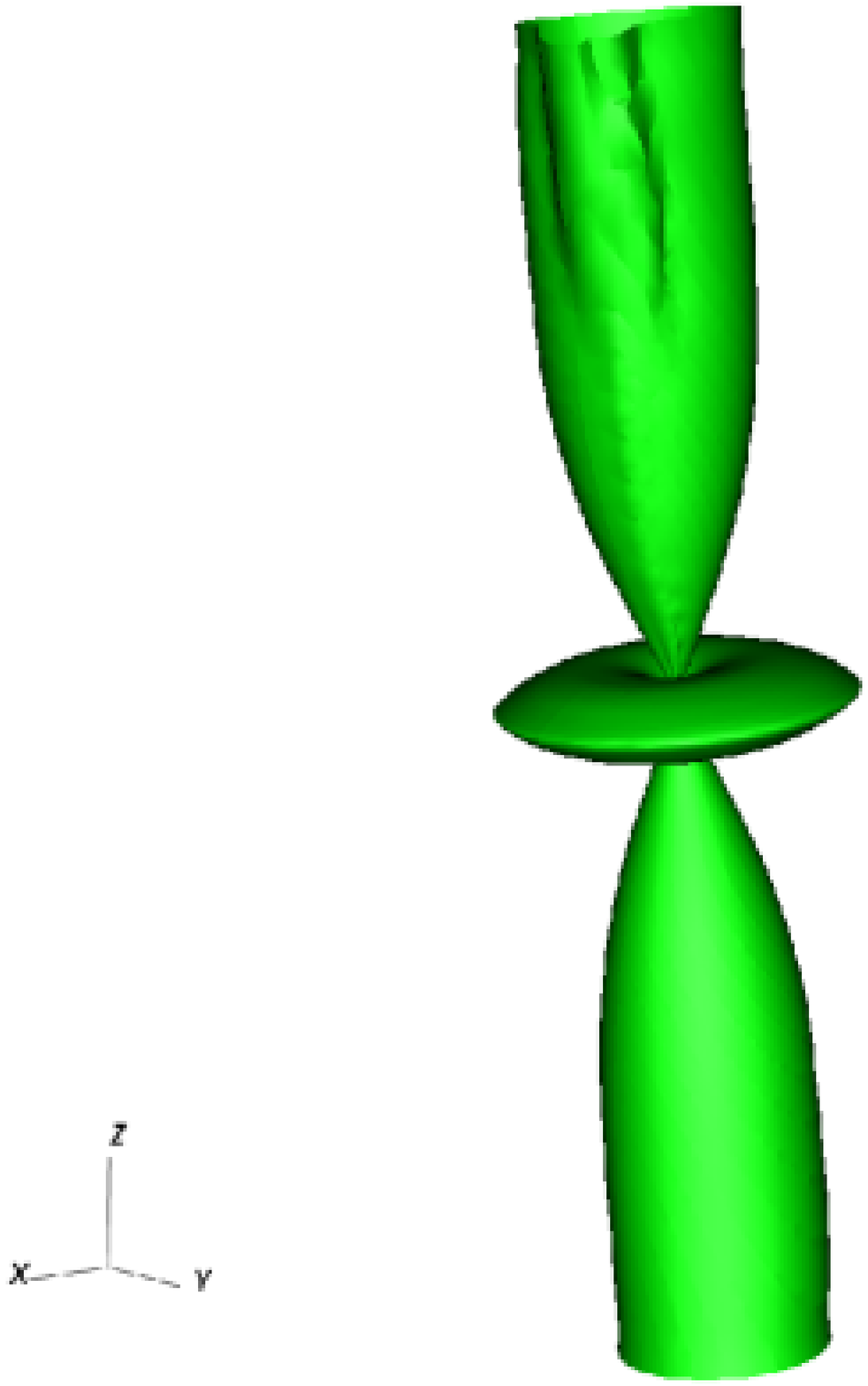}\tabularnewline
\end{tabular}

\end{center}

\caption{As in Fig.~\ref{fig:3-d-plots}, but for Models~III and IV. }

\label{fig:3-d-plots-2}
\end{figure*}


\subsection{Dependency on the disk tilt angle $\Theta$}

\label{sub:Dependency-on-beta}

We now examine the model dependency on disk tilt angle ($\Theta$)
while keeping all other parameters fixed. The results from Models~I,
II and III (c.f., ~Tab~\ref{tab:Model-Summary}), which use $\Theta=0^{\circ}$,
$5^{\circ}$ and $15^{\circ}$ respectively, are compared for this
purpose. Note that the observations suggest that a large fraction
of AGN have rather small i.e., $\Theta<\sim15^{o}$ (e.g.,~\citealt{Lu:2005}). 

Figure~\ref{fig:Density-and-velocity} shows the slices of the density
and velocity fields (on the $y=0$ plane) from snapshots of our four
simulations. The snapshots are chosen at the time when the models
reached a (semi-)steady state for Models~I and II. As we will see
later, the flow never reaches steady state in Model~III; therefore,
we chose the snapshot of the model at the time when the flow pattern
is a typical of a whole simulation time sequence. While accretion occurs
mainly on the equatorial plane ($z=0$) for Model~I, it occurs in
a inclined plane with a pitch angle (a angle between the equatorial
plane and the accretion plane) similar to the disk inclination angle,
for Models~II and III. In the precessing disk models (II and III),
the deviation from the axisymmetric is clearly seen in both density
distribution of gas and the shapes of the Mach number contours. Corresponding
3-D density and Mach number contour surfaces of these models are also
shown in Figs.~\ref{fig:3-d-plots} and \ref{fig:3-d-plots-2}. The
morphology of the density distribution seen in Figure~\ref{fig:Density-and-velocity}
resemble that of the Z-shaped (for Model~II) and the S-shaped (for
Model~III) radio jets (e.g.~\citealt{Condon:1984}; \citealt{Hunstead:1984};
\citealt{Tremblay:2006}) although in different scales. Obviously,
the difference between the Z- and S- shapes are simply due to the
difference in the viewing angles. Unlike the MHD precessing jet models,
the bending structures of the density distributions seen here are
shaped by the geometry of the sonic surfaces. When accreting material
from the outer boundary encounters the relatively low density but
high-speed outflowing gas launched by the radiation force from the
inner part, the gas becomes compressed, and forms higher density regions.
The flows in the bending density structure itself are rather complex
(especially in Model~III), but the direction of the flow becomes
outward (in radial direction) as they approach the sonic surface (excluding
the one shaped like a disk formed by the \emph{accreting} gas in the
inner region). Relatively large curvatures of the flows seen in both
density and the Mach number contours of Models~II and III can be
also understood from the fact the precession period used in these
models ($P=16000\,\mathrm{yr}$) is comparable to the gas free fall
time ($t_{\mathrm{ff}}=7000\,\mathrm{yr}$, c.f., \S~\ref{sub:Reference-Values}).
The curvatures or the {}``twists'' of the weakly collimated bipolar
flows can be clearly seen in the 3-D representation of these models
in Figs.~\ref{fig:3-d-plots} and \ref{fig:3-d-plots-2}. 


%
\begin{figure*}
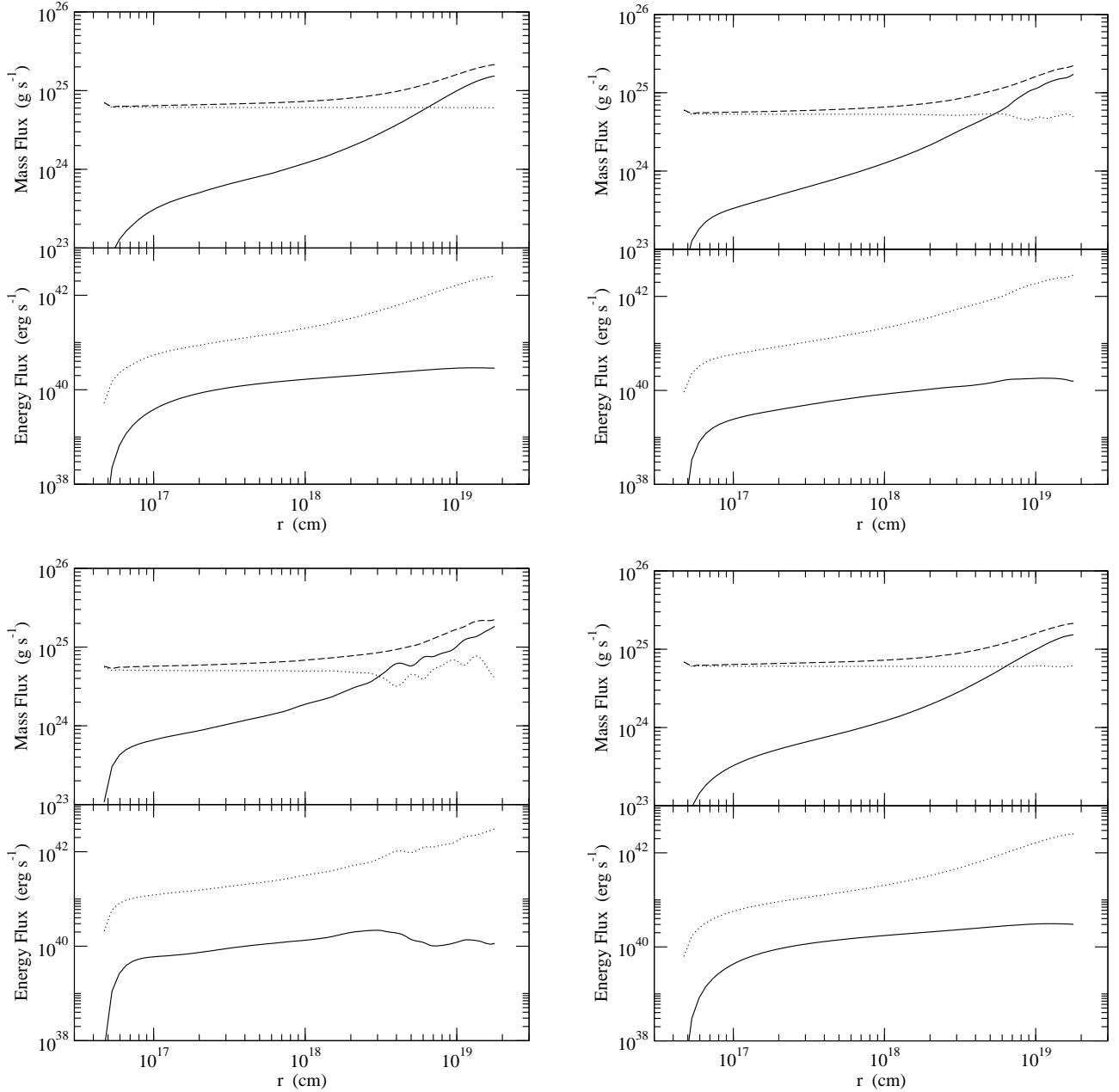

\begin{center}

\begin{tabular}{cc}
\includegraphics[clip,width=0.45\textwidth]{f5a.eps}&
\hspace{0.5cm}\includegraphics[clip,width=0.45\textwidth]{f5b.eps}\vspace{0.25cm}\tabularnewline
\includegraphics[clip,width=0.45\textwidth]{f5c.eps}&
\hspace{0.5cm}\includegraphics[clip,width=0.45\textwidth]{f5d.eps}\tabularnewline
\end{tabular}

\end{center}

\caption{Comparison of mass flux and energy flux as a function of radius for
Models~I (upper-left), II (upper-right), III (lower-left) and IV
(lower-right). The plot for each model is subdivided into two panels:
top (mass flux) and bottom (energy flux). In the mass flux plots,
the inflow (dash: $\dot{M}_{\mathrm{in}}$), outflow (solid: $\dot{M}_{\mathrm{o}}$)
and net (dot: $\dot{M}_{\mathrm{net}}$) mass fluxes, as defined in
equation~(\ref{eq:mdot2}), are separately plotted as a function
of radial distance from the center. The absolute values of $\dot{M}_{\mathrm{in}}$
and $\dot{M}_{\mathrm{net}}$ are plotted here since they are negative
at all radii. In the energy flux plots, the kinetic energy (solid)
and the thermal energy (dot), defined in eqs.~(\ref{eq:power-kinetic})
and (\ref{eq:power-thermal}), are shown. Note that the times slices
of the model simulations used here to computed the fluxes are same
as those in Figure~\ref{fig:Density-and-velocity}. }

\label{fig:mass-engery-flux}
\end{figure*}


We compute the mass fluxes as a function of radius for a quantitative 
comparison of the characteristics of the flows in the models. Following
\citet{Proga:2007}, the net mass flux ($\dot{M}_{\mathrm{net}}$),
the inflow mass flux ($\dot{M}_{\mathrm{in}}$) and the outflow mass
flux ($\dot{M}_{\mathrm{out}}$) can be computed from 
\begin{align}
  \dot{M}\left(r\right) & =\oint_{s}\rho\,\boldsymbol{v\,\cdot}d\boldsymbol{a}\label{eq:mdot}\\
                        & =r^{2}\oint_{4\pi}\rho v_{r}\, d\Omega\label{eq:mdot2}
\end{align}
where $v_{r}$ is the radial component of velocity $\boldsymbol{v}$.
In the equation above, $\dot{M}=\dot{M}_{\mathrm{net}}$ if all $v_{r}$
are included. Similarly, $\dot{M}=\dot{M}_{\mathrm{in}}$ for $v_{r}<0$
and $\dot{M}=\dot{M}_{\mathrm{out}}$ for $v_{r}>0$. Also, $d\boldsymbol{a}=\boldsymbol{\hat{r}}\, r^{2}\sin\theta\, d\theta\, d\phi$
and $d\Omega=\sin\theta\, d\theta\, d\phi$. Similarly we further
define the \emph{outflow} power in the form of kinetic energy ($P_{k}$)
and that in the thermal energy ($P_{\mathrm{th}}$) as functions of
radius i.e., 
\begin{align}
  P_{k}\left(r\right) & =r^{2}\oint_{4\pi}\rho v_{r}^{3}\,
  d\Omega
  \label{eq:power-kinetic}
\end{align}
and
\begin{align}
  P_{\mathrm{th}}\left(r\right) & =r^{2}\oint_{4\pi}uv_{r}\,
  d\Omega\,.
  \label{eq:power-thermal}
\end{align}
where $v_{r}>0$. 

The resulting mass fluxes and the outflow powers of the models are
summarized in Figure~\ref{fig:mass-engery-flux}. In all cases, the
mass inflow flux exceeds the mass outflow rate at all radii. For Models~I
and II, the net mass fluxes ($\dot{M}_{\mathrm{net}}$) are almost
constant at all radii, indicating that the flows in these models are
steady. Despite the presence of the disk precession in Model~II,
the flow becomes steady. The density distribution and the velocity
field become almost constant in the coordinate system co-rotating
with the disk precession period. On the other hand, $\dot{M}_{\mathrm{net}}$
for Model~III does not remain constant as $r$ becomes larger ($r>10^{18}\,\mathrm{cm}$)
because of the unsteady nature of the flow (c.f., Figs.~\ref{fig:Density-and-velocity}
and \ref{fig:3-d-plots-2}). As the disk tilt angle $\Theta$ increases,
the direction of the outflows, which are normally in polar directions
($\pm z$ directions) with an absence of the disk tilt, moves toward
the equatorial plane (the $x$-$z$ plane) where the flow is predominantly
inward. This opposite flows makes it harder for the outflowing gas
to reach the outer boundary. Further, since the disk is precessing,
the direction of the outflow is constantly changing. This results
in continuous collisions between the inflowing and outflowing gas
especially for a larger $\Theta$ model. The net mass fluxes at the
inner boundary $\dot{M}_{\mathrm{net}}\left(r_{\mathrm{i}}\right)$
are $-0.6$, $-0.6$ and $-0.4\times10^{25}\,\mathrm{g\, s^{-1}}$
(or equivalently $-0.10$, $-0.1$0 and $-0.06$~$\MsunPerYear$)
for Models~I, II and III respectively (Tab.~\ref{tab:Model-Summary}),
indicating the net mass flux inward (negative signs indicate inflow)
decreases slightly, but not significantly as the disk tilt angle $\Theta$
increases. 

The ratios of the total mass outflow flux to the total mass inflow
at the outer boundary ($\mu=\left|\dot{M}_{\mathrm{out}}/\dot{M}_{\mathrm{in}}\right|$)
are $0.7$3, $0.74$, $0.82$  for Models I, II, and III (see also
Tab.~\ref{tab:Model-Summary}). These values indicates that the high
efficiency of the outflow by the radiation pressure even for a modest
Eddington number used here i.e., $\Gamma=0.6$. This conversion efficiency
$\mu$ (from the outflow to inflow) is about the same for Models~I
and II, but it slightly ($\sim12$\%) increases for Model~III which
has the highest disk tilt angle. Overall characteristics of the mass-flux
curves as a function radius for Models~I and II are also very similar
to each other. The curves for Model~III are also similar to those
of Models~I and II; however, they differ in the outer radii ($r<\sim10^{18}\mathrm{cm}$),
mainly because of the unsteady nature of the flow in this model. 

The maximum speed of the outflow in the radial direction
$v_{r}^{max}\left(r_{\mathrm{\mathrm{o}}}\right)$ 
decreases as $\Theta$ increases (Tab.~\ref{tab:Model-Summary}). The
reduction in the speed is very significant ($\Delta v_{r}=-860\,\kmps$)
as $\Theta$ increase from $0^{\circ}$ to $5^{\circ}$ while the change
is relatively small ($\Delta v_{r}=-20\,\kmps$) as $\Theta$ changes
from $5^{\circ}$ to $15^{\circ}$. 

Figure~\ref{fig:mass-engery-flux} also shows the outflow powers
($P_{k}$ and $P_{\mathrm{th}}$) of the models as a function of radius,
as defined in eqs.~(\ref{eq:power-kinetic}) and (\ref{eq:power-thermal}).
As for the mass flux curves in the same figure, the dependency of
the energy flux curves on radius for Models~I, II and III are very
similar to each others. A small but noticeable deviations of the curves
for Model~III from those for Models~I and II are seen at the large
radii ($r>3\times10^{18}\,\mathrm{cm}$). The figure shows that in
all three models, the outflow power is dominated by thermal process
($P_{\mathrm{th}}\approx10-100\, P_{k}$ at all radii). This can be
explained by the high temperatures of the gas ($T\approx T_{C}=2.0\times10^{7}\,\Kelvin$
) in the computational domains caused by the (almost) isothermal equation
of state and the temperature ($2.0\times10^{7}\,\Kelvin$) fixed at
the outer boundary. The kinetic powers or the radiation forces are
not as significant as the pressure gradient force in these models;
however, their importance cannot be ignored since they {}``shape''
the geometry of the outflow as they strongly depend on the polar angle
position of a point in the computational field. We also note that
as $\Theta$ increases, the kinetic power at the outer boundary $P_{k}\left(r_{\mathrm{o}}\right)$
decreases significantly e.g.~$P_{k}\left(r_{\mathrm{o}}\right)$
of Model~III is three times smaller than that of the axisymmetric
model, Model~I (see Table~\ref{tab:Model-Summary}).

\begin{figure}[h]
\begin{center}

\includegraphics[clip,width=0.45\textwidth]{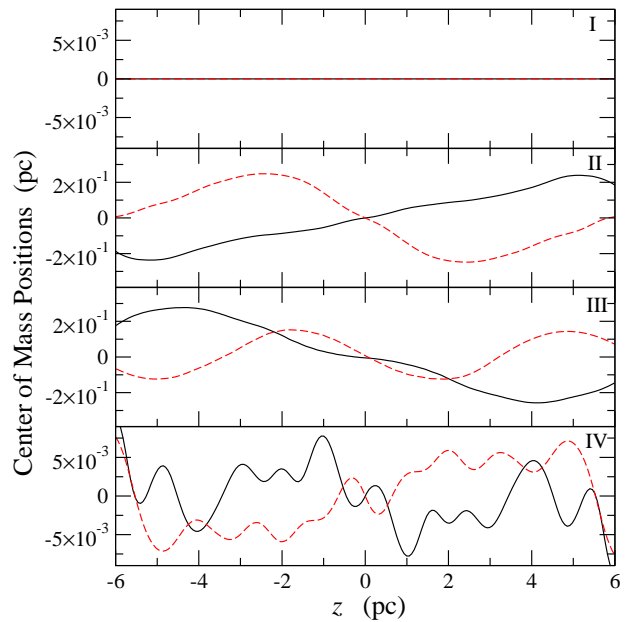}

\end{center}

\caption{Positions of the center of mass of the gas on the planes perpendicular
to the $z$-axis for Models I, II, III and IV from the top bottom.
The locations of the center of mass, $x_{c}$ (sold) and $y_{c}$
(dash) as defined in eqs.~(\ref{eq:cm-x-comp}) and (\ref{eq:cm-y-comp}),
are computed as a function of $z$. All models except Model~I show
a clear sign of deviations from axisymmetry; however the displacement
of the center of mass remains fairly small ($\left|x_{c}\right|<\sim0.3\,\mathrm{pc}$
and $\left|y_{c}\right|<\sim0.3\,\mathrm{pc}$), at all $z$ locations,
compared to the size of outermost radius $r_{\mathrm{o}}$ ($7.1$~pc)
of the computational domain. The patterns in $x_{c}$ and $y_{c}$
curves for Models~II and III indicate that the flow density structures
are helical.}

\label{fig:Center-of-mass}
\end{figure}

Next we examine the degree of non-axisymmetry in Model~I, II and
III, and seek for any obvious dependency on $\Theta$. For this purpose,
we compute the center of mass (CM) of the gas on the planes perpendicular
to the $z$-axis (as a function of $z$) i.e., $x_{c}\left(z\right)$
and $y_{c}\left(z\right)$ which are defined as 
\begin{equation}
  x_{c}\left(z\right)=\frac{\int_{-r_{\mathrm{o}}}^{r_{\mathrm{o}}}\int_{-a}^{a}\int_{-\infty}^{\infty}x\,\delta\left(z-z'\right)\rho\left(x,y,z'\right)dz'\, dx\, dy}{m\left(z\right)}
  \label{eq:cm-x-comp}
\end{equation}
and 
\begin{equation}
  y_{c}\left(z\right)=\frac{\int_{-r_{\mathrm{o}}}^{r_{\mathrm{o}}}\int_{-a}^{a}\int_{-\infty}^{\infty}y\,\delta\left(z-z'\right)\rho\left(x,y,z'\right)dz'\, dx\, dy}{m\left(z\right)}
  \label{eq:cm-y-comp}
\end{equation}
where 
\begin{equation}
  m\left(z\right)=\int_{-r_{\mathrm{o}}}^{r_{\mathrm{o}}}\int_{-a}^{a}\int_{-\infty}^{\infty}\delta\left(z-z'\right)\rho\left(x,y,z'\right)dz'\, dx\, dy\,\,,
  \label{eq:mass-on-plane}
\end{equation}
and $a\left(y,z\right)=$ $\left(r_{\mathrm{o}}^{2}-z^{2}-y^{2}\right)^{1/2}$.
The results are shown in Figure~\ref{fig:Center-of-mass}. As expected,
the CM position remains constant and on the $z$-axis ($x_{c}\left(z\right)=0$
and $y_{c}\left(z\right)=0$) for Model~I, as this is an axisymmetric
model. For both Models~II and III, the maximum amount of deviations
for each component of the CM ($\left|x_{c}\right|$and $\left|y_{c}\right|$)
is about $0.3$~pc which is relatively small compared to the outer
boundary radius ($r_{\mathrm{o}}=7.1\,\mathrm{pc}$). The $x_{c}$
and $y_{c}$ curves are anti-symmetric about the $z=0$ position since
our model accretion disk hence the radiation force is symmetric about
the origin of the coordinate system. The plot also shows that the
positions of the maxima and minima in the $x_{c}$ and $y_{c}$ curves
do not coincide, but they are rather shifted in both $+z$ and $-z$
directions. This clearly demonstrates a helical or twisting nature
of the flows, as one can also simply see it in the 3-D density and
Mach number contour plots in Figs.~\ref{fig:3-d-plots} and \ref{fig:3-d-plots-2}. 

To summarize, as the tilt angle of the disk precession $\Theta$ increases,
reductions of the maximum outflow velocity ($v_{r}$) and the kinetic
outflow power ($P_{k}$) at the outer boundary $r_{\mathrm{o}}$ occur,
as a consequence of the stronger interactions between the outflowing
and inflowing gas of as $\Theta$ increases. The net mass inflow flux
($\dot{M}_{\mathrm{net}}$) at the inner boundary does not strongly
depend on $\Theta$. The thermal outflow energy power dominates the
kinetic outflow power in our models here because of the high temperature
of set at the outer boundary and because the gas is (almost) isothermal.
The flows of Models~II and III show helical structures; however,
the radius of the helices (base on the CM positions along the z-axis)
does not change greatly as $\Theta$ increases from $5^{\circ}$ to
$15^{\circ}$. 


%


\subsection{Dependency on the disk precession period $P$}

\label{sub:Dependency-on-P}

We now examine the dependency of the model on the disk precession
period ($P$). We vary the value of $P$ while fixing the disk tilt
angle to $\Theta=5^{\circ}$. For this purposed, we compare Models~I,
II and IV as summarized in Table.~\ref{tab:Model-Summary}. The precession
periods $P$ are $\infty$, $1.6\times10^{4}$ and $1.6\times10^{5}$~yr
respectively for Models~I, II and IV. In the units of the free-fall
time ($t_{\mathrm{ff}}=7.0\times10^{3}$~yr ) from the Bondi radius
(\S~\ref{sub:Reference-Values}), they are $\infty$, $2.3$ and
$23$ respectively. Note that the observations suggest that typical
values of jet precession period are $P=10^{4}$--$10^{6}$~yr (c.f.,
Tab.1 in \citealt{Lu:2005}). 

Figure~\ref{fig:Density-and-velocity} shows that similarities between
Model~IV and Model~I in their morphology of the density distribution
and Mach number surfaces. At a given time, the flow in Model~IV is almost axisymmetric,
and the symmetry axis is tilted also by $\Theta=5^{\circ}$ from the
$z$-axis. This is caused by the relatively long precession period
for Model~IV compared to the dynamical time scale or the gas free-fall
time scale $t_{\mathrm{ff}}$. The curvature or helical motion of
the gas is not significant, and it does not greatly affect the overall
morphology of the flow, except for the outermost part of the flow
where the flow is slightly turbulent due to the shear of the slowly
precessing flow and the outer boundary. This can be clearly seen in
the 3-D plots Figure~\ref{fig:3-d-plots-2}. As the precession period
$P$ becomes shorter and comparable to $t_{\mathrm{ff}}$ (as in Model~II),
the flow shows more curvature and the helical structures. 

The mass flux curves (c.f., eq.~{[}\ref{eq:mdot2}]) for Models~IV
in Figure~\ref{fig:mass-engery-flux} also show that nature of the flows between
Models~I and IV are very similar to each other. Overall characteristics
of the curves are also similar to that of Model~II. In fact, the
net mass flux $\dot{M}_{\mathrm{net}},$ the inflow mass flux $\dot{M}_{\mathrm{in}}$
and the outflow mass flux $\dot{M}_{\mathrm{out}}$ at the outer boundary
of Model~IV are identical to those of Model~I (see Table~\ref{tab:Model-Summary}).
Also note that Models~I, II and IV all have same $\dot{M}_{\mathrm{in}}$
value at the inner boundary, i.e., the mass inflow rate across the
inner boundary in insensitive to the change in the precession period
for $\Theta=5^{\circ}$. 

The similarity between and Models~I and IV can be also seen in the
outflow powers, $P_{k}$ and $P_{\mathrm{th}}$. Figure~\ref{fig:mass-engery-flux}
shows $P_{k}$ and $P_{\mathrm{th}}$ as a function of radius for
Model~IV are almost identical to those of Model~I. The $P_{k}$
and $P_{\mathrm{th}}$ values at outer boundary are indeed identical
(Table~\ref{tab:Model-Summary}). A slight increase in the maximum
outflow velocity at the outer boundary $v_{r}^{\mathrm{max}}(r_{\mathrm{o}})$
is observed for Model~IV, compared to Model~I. The kinetic outflow
power $P_{k}$ and the maximum outflow velocity at the outer boundary
$v_{r}^{\mathrm{max}}(r_{\mathrm{o}})$ decreases as the precession
period become comparable to $t_{\mathrm{ff}}$. 

The CM positions  $x_{c}$ and $y_{c}$ as a function of $z$ (see
\S~\ref{sub:Dependency-on-beta}) for Model~IV is shown in Figure~\ref{fig:Center-of-mass}.
Compared to Model~II, the maximum displacement of the CM is about
40 times smaller in Model~IV i.e.~$\left|x_{c}\right|<\sim0.075\,\mathrm{pc}$
and $\left|y_{c}\right|<\sim0.075\,\mathrm{pc}$. The $x_{c}$ curve
for Model~IV shows a rather complex pattern compared to that in Model~II.
This and the visual inspection of the density and the Mach number
contour surfaces in Figure~\ref{fig:3-d-plots-2}, indicates that
the bipolar outflow flows are slightly twisted, but does not have
clear helical structure.

\subsection{Time Evolution of Mass Accretion/Outflow Rates and Angular Momentum }

\label{sub:Time-dependet-behavours}


%
\begin{figure*}
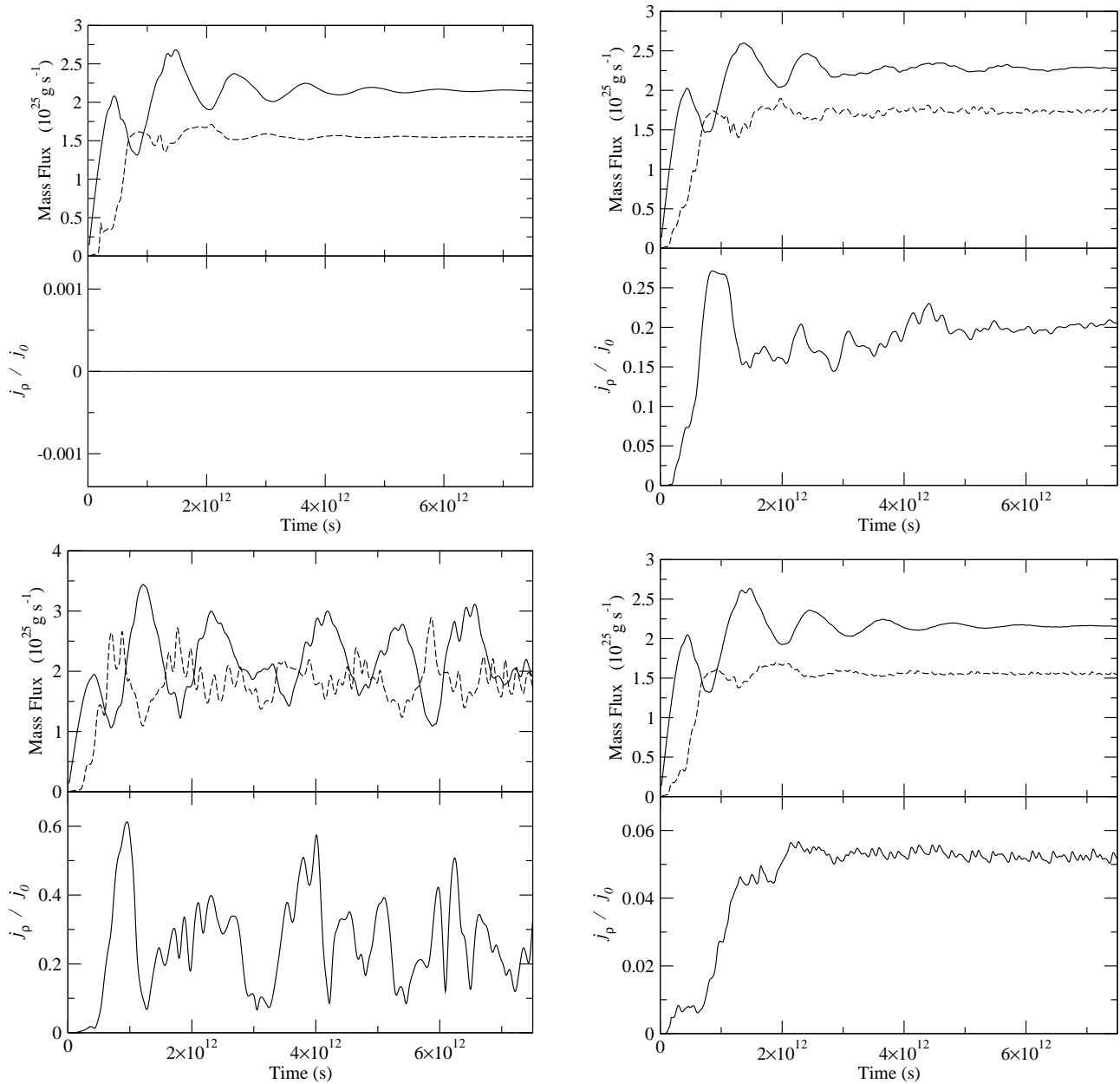

\begin{center}

\begin{tabular}{cc}
\includegraphics[clip,width=0.45\textwidth]{f7a.eps}&
\hspace{0.5cm}\includegraphics[clip,width=0.45\textwidth]{f7b.eps}\tabularnewline
\includegraphics[clip,width=0.45\textwidth]{f7c.eps}&
\hspace{0.5cm}\includegraphics[clip,width=0.45\textwidth]{f7d.eps}\tabularnewline
\end{tabular}

\end{center}

\caption{The mass flow rates across the outer boundary and the density-weighted
mean specific angular momentum $j_{\rho}$ (c.f.~eqs.~[\ref{eq:mdot2}]
and [\ref{eq:def_rho_mean_j}]) plotted as a function of time for Models~I
(upper-left), II (upper-right), III (lower-left), and IV (lower-right).
For each model, the plot is subdivided into two panels: mass-inflow/outflow
rates (top panels) and $j_{\rho}$ (bottom panels). In the top panels,
the mass-inflow rate at the outer boundary (solid), and the mass-outflow
rate at the outer boundary (dash) are shown separately. The values
of $j_{\rho}$ are in the units of $j_{0}$ which is defined as the
specific angular momentum of the gas in Keplerian orbit at the inner
boundary ($r=r_{\mathrm{in}}$). Note that the precession period used
here are $\infty$, $5.0\times10^{11}\,\mathrm{s}$, $5.0\times10^{11}\,\mathrm{s}$
and $5.0\times10^{12}\,\mathrm{s}$ for Models I, II, III and IV respectively.}

\label{fig:mdot-j-evolution}
\end{figure*}


Next, we examine the variability or steadiness of the flows in each
model by monitoring the mass fluxes at the outer boundary as in equation~(\ref{eq:mdot2})
and the angular momentum of the system as a function of time. For
the latter, we compute the density-weighted mean specific angular
momentum $\boldsymbol{j}_{\rho}$ of the systems defined as: 
\begin{equation}
  \boldsymbol{j}_{\rho}=\frac{\int_{V}\rho\,\left(\boldsymbol{r}\times\boldsymbol{v}\right)\, dV}{\int_{V}\rho\, dV}
  \label{eq:def_rho_mean_j}
\end{equation}
where the denominator is simply the total mass of the gas in the computational
domain. Note that the radiation force (eq.~[\ref{eq:rad-force-final}])
and the gravitational force are in radial direction only. Consequently,
they do not exert torque onto the system; hence, they do not contribute
to the change in the angular momentum of the system directly. The
system can gain the angular momentum in the following way. In our
models, the strength of the
disk radiation field depends on the angle measured from the disk
normal (c.f., eq.~[\ref{eq:disk-flux}]). 
This causes gas pressure gradients in azimuthal direction, and  
contributes to the angular momentum of gas locally,  forming
vorticity. The precession of radiation field hence the precessing
outflow will cause the gas with preferred sign of vorticity to escape
from the outer boundary, resulting a change in the net angular momentum
of the gas in the computational domain. 

Figure~\ref{fig:mdot-j-evolution} shows $\dot{M}_{\mathrm{in}}\left(r_{o}\right)$,
$\dot{M}_{\mathrm{out}}\left(r_{o}\right)$ and $j_{\rho}$ for Models~I,
II, III and IV as a function of time. For Models I, II and IV, both
the mass fluxes and $j_{\rho}$ reach to asymptotic values by $t\approx7\times10^{12}\,\mathrm{s}$. Small
oscillations of $j_{\rho}$ around the asymptotic values are seen
for Models~II and IV. On the other hand, the mass fluxes of Model~III
has much larger amplitudes of the oscillations around an asymptotic
value. By visual inspections, their oscillations do not seem have
a clear periodicity associated with them. We performed the Lomb-Scargle
periodgram analysis (e.g.~\citealt{Horne:1986}; \citealt{Press:1992})
on the $\dot{M}_{\mathrm{in}}\left(r_{o}\right)$, $\dot{M}_{\mathrm{out}}\left(r_{o}\right)$
and $j_{\rho}$ curves for Model~III. Only $\dot{M}_{\mathrm{out}}\left(r_{o}\right)$
shows a relatively strong signal at $P_{\mathrm{LS}}=1.36\times10^{12}\,\mathrm{s}$
which is about $2.7$ times longer than the precession period of Model~III.
On the other hand $\dot{M}_{\mathrm{in}}\left(r_{o}\right)$ and $j_{\rho}$
curves do not have any obvious period associated with them, but they
are rather stochastic. 

As mentioned in \S~\ref{sub:Dependency-on-beta}, as the disk tilt
angle $\Theta$ increases the direction of the outflows, which normally
exists in polar directions with an absence of the disk tilt, moves
toward the equatorial plane (the x-$z$ plane) where the flow is predominantly
inward. In addition, the precession of the disk causes the direction
of the outflow to change constantly; hence, causing constant creation
of the shock between the inflowing and the outflowing gas. This leads
into a very unstable flow of the gas at all time for a model with
a larger $\Theta$ e.g., Model~III with $\Theta=15^{\circ}$. The flow,
of course, can be stabilized if the precession period is increased
to a value much larger than the free-fall time $t_{\mathrm{ff}}$. 

The amount of the (density-weighted) mean specific angular momentum
deposited to the gas by the precessing disk (measured by $j_{\rho}$)
is largest in Models~II and III (see Table~\ref{tab:Model-Summary}),
and that in Model~IV is about $4$ times less than those of Models~II
and III. For all models, a time-averaged value (by using the last
$2\times10^{12}\,\mathrm{s}$ of the simulation) of $j_{\rho}$ is
used in Table~\ref{tab:Model-Summary}. It seems that the faster
the disk precesses, the lager the amount of angular momentum transferred
to the environment; however, this trend does not continues, as we
increase the disk precession speed even faster. Although not shown
here, a model with exactly the same set of parameters as in Model~II,
but with $P=1600\,\mathrm{yr}$ (10 times faster rotation), showed
that the the value of $j_{\rho}$ decrease to $\sim0.01$, which is
even smaller than Model~IV (with $P=160000\,\mathrm{yr}$). This
indicates that the amount of angular momentum deposited to the gas
depends on how close the precession period to the dynamical time scale
of the flow. 

In principle, it is possible to model the change in the angular momentum
of the accretion disk itself through the transfer of angular momentum
from the environment, we ignored this effect for simplicity (and this
is also our model limitation). To model the interaction of the disk
angular momentum and the angular momentum of the surrounding gas properly,
we need to model the dynamics of the gas in the accretion disk itself
as well as the dynamic of the gas which is much larger scale as in
our models here. This is computationally challenging with our current
code since we have to resolve the length scale of the innermost part
of the accretion disk ($\sim10^{-5}$~pc) to the large scale outflow/inflow
gas ($\sim10$~pc).

\section{Conclusions}

\label{sec:Conclusions}

We have studied the dynamics of the gas under the influences of the
gravity of SMBH and the radiation force from the luminous accretion
disk around the SMBH. The rotational axis of the disk was assumed
to be tilted with respect to the symmetry axis with a given angle
$\Theta$ and a precession period $P$ (c.f., Figure~\ref{fig:model-config}).
We have investigated the dependency of the flow morphology, mass accretion/outflow
rates, angular momentum of the flows for different combinations of
$\Theta$ and $P$. This is a natural extension of similar but more
comprehensive 2-D radiation hydrodynamics models of AGN outflow models
by ~\citet{Proga:2000}, \citet{Proga:2007}. As this is our first
attempt for modeling such gas dynamics in full 3-D, we have used a
reduced set of physical models described in \citet{Proga:2007} i.e.,
the radiation force due to line and dust scattering/absorption, and
the radiative cooling/heating are omitted. In the following, we summarize
our main findings through this investigation. 

(1)~Our assumption of the adiabatic index ($\gamma=1.01$) keeps
the mean temperature of the gas in the computational domain relatively
high ($\sim2\times10^{7}\,\Kelvin$) which is essentially determined
by the outer boundary condition. For our axisymmetric model (Model~I:
Figs.~\ref{fig:Density-and-velocity} and \ref{fig:3-d-plots}),
this results in the flow morphology very similar to the model with
a relatively high X-ray heating (see Run~A in \citealt{Proga:2007})
in which the line force is inefficient because of the high gas temperature
and hence the high ionization state of the gas. 

(2)~Although in different scales, we were able to reproduced the
Z- or S- shaped density morphology of the gas outflows (Fig.~\ref{fig:Density-and-velocity})
which are often seen in the radio observations of AGN (e.g.~\citealt{Florido:1990};
\citealt{Hutchings:1988}). The bending structure seen here are shaped
by the shape of the sonic surfaces. When accreting material from the
outer boundary encounters the relatively low density but high speed
outflowing gas launched by the radiation force from the inner part,
the gas becomes compressed, and forms higher density regions. 

(3)~As the tilt angle of the disk precession $\Theta$ increases,
the reduction of the maximum outflow velocity ($v_{r}$) and the kinetic
outflow power $P_{k}$ at the outer boundary $r_{\mathrm{o}}$ decrease
as a consequence of the stronger interactions between the outflowing
and inflowing gas (Tab.~\ref{tab:Model-Summary}). The net mass inflow
rate ($\dot{M}_{\mathrm{net}}$) at the inner boundary does not change
significantly with increasing $\Theta$. 

(4)~A relatively high efficiency of the outflow ($\mu=\dot{M}_{\mathrm{out}}/\dot{M}_{\mathrm{in}}$)
by the radiation pressure were observed in our models ($70$--$80$~\%;
see also Tab.~\ref{tab:Model-Summary}) for a Eddington number ($\Gamma$)
of $0.6$ here. The conversion efficiency $\mu$ (from the outflow
to inflow) is about the same for Models~I and II, but it slightly
($\sim12$\%) increased for Model~III which has the highest disk
tilt angle. 

(5)~The thermal outflow energy power dominates the kinetic outflow
power (Fig.~\ref{fig:mass-engery-flux}) in the models presented
here because of the high temperature of the flow (as mentioned above). 

(6)~The flows of Models~II and III show helical structures (c.f.,
Figs.~\ref{fig:3-d-plots} and \ref{fig:3-d-plots-2}); however,
the radius of the helices does not change as $\Theta$ increases from
$5^{\circ}$ to $15^{\circ}$, based on the locations of the center
of mass (Figure~\ref{fig:Center-of-mass}) of the planes perpendicular
to the symmetry axis (the $z$-axis in Fig.~\ref{fig:model-config}).
We leave for a future investigation to test whether these trends continue
as $\Theta$ becomes larger than $15^{\circ}$. 

(7)~The characteristics of the flows are closely related to a combination
of $P$ and $\Theta$, but not to $P$ and $\Theta$ individually. Even
with a relatively large disk tilt angle $\Theta$, if the precession
period is much larger than the dynamical time scale of a system, the
flow geometry obviously becomes almost axisymmetric (c.f.~Model~IV
in Figs.~\ref{fig:Density-and-velocity} and \ref{fig:3-d-plots-2}). 

(8)~The gas dynamics of a model with a relative large disk tilt angle
($\Theta=15^{\circ}$) with a precession period comparable to the gas
free-fall time ($t_{\mathrm{ff}}$) of the system (e.g.,~Model~III)
does not reach a steady state because the outflows driven by the luminous
accretion disk constantly collides with the inflowing/accreting gas
as the disk precesses hence as the outflow direction changes. 

(9)~The amount of the density-weight mean specific angular momentum
($j_{\rho}$) deposited by the precessing disk is largest for Models~II
and III (Tab.~\ref{tab:Model-Summary} and Fig.~\ref{fig:mdot-j-evolution})
which have the precession period comparable to $t_{\mathrm{ff}}$. 

The models represented here are mainly for exploratory purpose --
to examine the basic model dependencies on $\Theta$ and $P$ -- with
a relatively simple set of physics but in full 3-D. In the follow-up
paper, we will improve our model by including the physics omitted
here (the line scattering/absorption, dust scattering/absorption,
and the radiative cooling/heating) as in the 2-D models of e.g.,~\citet{Proga:2007}.

\acknowledgements{}

This work was supported by NASA through grant HST-AR-10680 from the
Space Telescope Science Institute, which is operated by the Association
of Universities for Research in Astronomy, Inc., under NASA contract
NAS5-26555. This work was also supported by the National Center for
Supercomputing Applications under AST070036N and utilized the Xeon
Linux Cluster, Tungsten. Authors are grateful for original developer
of ZEUS-MP for making the code publicly available. We thank Prof.~Jim
Pringle and Monika Mo\'{s}cibrodzka for the critical reading of the
manuscript, and comments. We also thank Agnieszka Janiuk for helpful
discussion and support. 


\end{document}